\documentclass[useAMS,usenatbib]{mnras}
\usepackage[T1]{fontenc}
\usepackage{ae,aecompl}
\usepackage{pdflscape}
\usepackage{graphicx}
\usepackage{longtable}
\usepackage{lscape}
\usepackage{amssymb}
\usepackage{endnotes} 
\usepackage{footnote}
\usepackage{subfigure}
\usepackage{amsmath}
\usepackage{txfonts}
\usepackage{natbib}
\usepackage{url}
\usepackage{breakurl}
\usepackage{helvet}
\usepackage[flushleft]{threeparttable}
\def\aj{{AJ}}
\def\araa{{ARA\&A}}
\def\apj{{ApJ}}
\def\apjl{{ApJ}}
\def\apjs{{ApJS}}
\def\aap{{A\&A}}
\def\aapr{{A\&A~Rev.}}
\def\aaps{{A\&AS}}
\def\mnras{{MNRAS}}
\def\pasp{{PASP}}

\def\HI{H{\small{I}}~}
\def\HII{H{\small{II}}~}
\def\HeII{He{\small{II}}~}

\title[Dwarf Wolf-Rayet galaxies]{Optical spectroscopy of star-forming regions in dwarf Wolf-Rayet galaxies}
\author[A. Paswan, A. Omar and S. Jaiswal]{A.~Paswan \thanks{E-mail:
p.abhishek@aries.res.in}, A. Omar \thanks{E-mail:
aomar@aries.res.in} and S. Jaiswal\\ Aryabhatta Research Institute of Observational Sciences, Manora Peak, Nainital 263002, India\\ Pt. Ravishankar Shukla University, Raipur, 492010, India}

\date{Accepted ------------, Received ------------; in original form ------------}
\pagerange{\pageref{firstpage}--\pageref{lastpage}} \pubyear{}
      
\begin{document}
\label{firstpage}
\pagerange{\pageref{firstpage}--\pageref{lastpage}} \pubyear{}
\maketitle

\begin{abstract}

We present here spatially-resolved optical spectroscopic observations of four nearby dwarf Wolf-Rayet (WR) galaxies. The ages of the most recent starburst events in these galaxies are found between 3 and 10 Myr. The gas-phase metallicities [12+log(O/H)] for the spatially-resolved star-forming regions are derived using several indicators. The star-forming regions within the galaxies are found chemically homogeneous within the uncertainties in the estimates. Nitrogen-enrichment as expected in the WR regions is not detected. This implies that metal-enrichment due to supernovae explosions in the most recent star-forming episode is not being detected here. It is suggested that the newly synthesized metals still reside in hot gas-phase. The metals from the previous episodes, cooled by now and well mixed across the whole extent of galaxies, are making galaxies chemically homogeneous with normal N/O ratio. These galaxies are residing in dense environments with galaxy density in the range of $8-80$ Mpc$^{-3}$.   

\end{abstract}



\section{Introduction}

Dwarf galaxies are ubiquitous in the local Universe. About $80-90$\% members of the local group \citep{1998ARA&A..36..435M,2001dge..conf...45G} and about 80\% of the known galaxies in the local volume \citep[$D$ $\leq$ 10 Mpc;][]{2004AJ....127.2031K} are classified as dwarf galaxies. Dwarf galaxies are generally defined based on their absolute magnitude M$_{B}$ $\gtrsim$ -16 mag \citep{1994ESOC...49....3T} or M$_{V}$ $\gtrsim$ -18 mag \citep{2001dge..conf...45G,2001slbt.work...79G}. Their space density in the Universe is about 40 times that of the brighter galaxies \citep{1992MNRAS.258..725S}. Dwarf galaxies usually have low stellar-mass ($\leq$ 10$^{10}$ M$_{\odot}$), low luminosity (M$_{B}$ $\gtrsim$ -18 mag), high gas content (M$_{HI}$ $\gtrsim$ 10$^{8}$ M$_{\odot}$) and low metallicity between $7.0 - 8.4$ \citep{1972ApJ...173...25S,1999ApJ...527..757I,2000A&ARv..10....1K,2004ApJS..153..429K,2008A&A...491..113P}. Dwarf galaxies play an important role in the formation and evolution of galaxies. In hierarchical models of galaxy growth, larger structures like giant spiral and massive elliptical galaxies form through mergers and accretion of smaller structures like dwarf galaxies \citep[e.g.,][]{1991ApJ...379...52W,1997MNRAS.286..795K,2013seg..book..555S,2014Natur.507..335A,2014ApJ...794..115D}. 

Dwarf galaxies are further classified as irregulars (dIs), dwarf ellipticals (dEs), dwarf spheroidals (dSphs), dwarf spirals (dSs) and blue compact dwarfs (BCDs) based on their optical appearances. Several possibilities of evolutionary connections between different types of dwarf galaxies have been proposed, however, this issue in not completely resolved \citep{1985ApJ...299..881T,1988MNRAS.233..553D,1991AJ....101...94D,1994MNRAS.269..176J,1996A&A...314...59P}. Among these sub-types, BCDs appear extremely blue due to recent starburst activity in a compact ($\textless$ 1 kpc) region \citep{1965ApJ...142.1293Z,1981ApJ...247..823T}. Since starburst activity in BCDs takes place within an underlying old stellar population, therefore, BCDs may not be considered young systems \citep[e.g.,][]{1986sfdg.conf...73L,1995A&A...303...41K,1996A&AS..120..207P,2002A&A...390..891B,2003A&A...410..481N,
2005ApJS..156..345G,2005RMxAC..24..223C,2013ApJ...764...44Z}. The very young ($\lesssim$ 10 Myr) stellar population dominated by O/B type stars in the star-bursting regions give blue colors to these galaxies. Their optical spectra are dominated by strong emission lines attributed to the ongoing star formation. BCDs were once considered equivalent to the primeval galaxies undergoing their first episode of star formation in the presence of nearly pristine interstellar medium \citep[ISM;][]{1970ApJ...162L.155S}. The issue as to whether BCDs exhibit old stellar populations or not has been widely explored for several years \citep[e.g.,][]{1970ApJ...162L.155S,1973ApJ...179..427S,1999ApJ...511..639I,2004ApJ...602..200I,2005ApJ...631L..45A}. Majority of BCDs have been found to have an underlying old stellar population with ages between $1-10$ Gyr. These studies also indicated that the starburst activities in BCDs do not last longer than about a few tens of Myr. The star formation in BCDs is inferred episodic with intense star formation activities separated by relatively long phase of quiescence \citep{1991ApJ...370...25T,1995A&A...303...41K,1999A&A...349..765M,2000ApJ...539..641T}. Very few \HI - selected dwarf galaxies in the quiescent phase without H$\alpha$ emission have been seen with non-zero star formation rate \citep[SFR;][]{2001AJ....121.2003V,2009ApJ...703..692L}, which implies that dwarf galaxies also maintain a low level of star formation activities over long periods.

BCDs show intense star-forming activity, fed by relatively large amount of gas \citep{1981ApJ...247..823T,1992MNRAS.258..334S,1998AJ....116.1186V}. Ultimately, the formation of young massive stars ($\textgreater$ $8 - 10$ M$_{\odot}$) and their subsequent evolutions cause the fresh metals (oxygen and other $\alpha$ elements) ejection into the ISM via stellar winds and supernova explosions. The released metals will be dispersed and mixed with the ISM via hydrodynamical process in timescales of a few 100 Myr \citep[e.g.,][]{1995A&A...294..432R,1996AJ....111.1641T}. This implies that the spatial distribution of the metal abundances in galaxies is a function of the recycling and mixing time scales of the ISM. The spatially-resolved abundance analysis in galaxies can therefore provide important insights about the chemical evolution of galaxies. These issues  have been addressed by studying the spatial distribution of optical emission line ratio and of certain elemental abundances in dwarf galaxies \citep[e.g.,][]{1996ApJ...471..211K,1997ApJ...477..679K,2006ApJ...642..813L,1989ApJ...341..722W}. 

Some studies have revealed spatial variations in the chemical compositions as measured from the gas-phase metallicity in different types of dwarf galaxies. For example, a shallow gradient ($\gtrsim$ 0.11 dex kpc$^{-1}$) in metallicity was inferred in SBS 0335-052E \citep{2006A&A...454..119P}. Inhomogeneous metallicity has been seen in extremely metal-poor galaxies, where it was noticed that the low metallicity regions are normally associated with intense star-forming regions \citep{2006A&A...454..119P,2009ApJ...690.1797I,2011ApJ...739...23L}. The existence of significantly large metallicity gradient or inhomogeneity within the galaxies is often understood in terms of a recent merger of two galaxies with different metallicities and tidal interactions \citep{2004A&A...428..425L,2004ApJS..153..243L,2009A&A...508..615L,2010A&A...517A..85L,2018MNRAS.473.4566P}. It may be noted here that typical metallicity gradients between -0.009 and -0.231 dex kpc$^{-1}$ are common in large spiral galaxies \citep{1994ApJ...420...87Z}. 

On the other hand, a good number of studies have also indicated that BCD galaxies have homogeneous chemical abundances \citep{1996ApJ...471..211K,2006A&A...454..119P,2008A&A...477..813K,2009ApJ...707.1676C,2009MNRAS.398..949P,
2011MNRAS.412..675P,2011MNRAS.414..272H,2013AdAst2013E..20L}. The chemical homogeneity in galaxies is explained as a consequence of starburst driven feedback that disperses and mixes the newly synthesized elements in the ISM through hydrodynamic processes \citep{1996AJ....111.1641T}. An issue related to metallicity is also the nitrogen-to-oxygen (N/O) ratio, which is observed to be increasing at high metallicities albeit with a large scatter. The production of nitrogen and oxygen and its subsequent maintenance in the gas-phase in the ISM is not completely understood. Massive stars produce small amounts of nitrogen in early phase of evolution, which is termed as the primary production of nitrogen \citep[]{1978MNRAS.185P..77E,1979A&A....78..200A,1999ApJ...511..639I}. Low and intermediate mass stars produce nitrogen and other elements heavily enriching ISM with a significant time lag from the primary production timescales. This latter process is often termed as the secondary production. The low metallicity regions (12+log(O/H) $\leq$ 7.8) with a constant N/O ratio around -1.6 is believed to be primarily due to primary production of nitrogen in massive stars. At high metallicity, a steep increase in N/O ratio is observed which is due to increased secondary production and partly also due to selective depletion of oxygen in dust grains \citep{2000ApJ...541..660H,2006A&A...448..955I,2008A&A...485..657B,2009MNRAS.398..949P,2010A&A...517A..85L,2015MNRAS.449..867B,2016MNRAS.458.3466V}. The mixing of recently produced metals with the surrounding ISM can also modify the observed N/O ratio. 

\begin{table*}
\centering
\caption{General properties of the galaxies in the present sample.}
\begin{tabular}{ccccccccc} \hline
Galaxy name & RA (J2000) & Dec (J2000) & Type & Distance & $V_{helio}$ & $M_{B}$ & Optical size & Other name\\
\hline
& [h:m:s]    &	[d:m:s] & & [Mpc] & [km s$^{-1}$] & [mag] & [arcmin $\times$ arcmin] &\\ \hline
IC 3521      & 12 34 39.5 & +07 09 37 & IBm  & 12.7 &  595 & -16.7 & 1.43 $\times$ 0.93 & UGC 7736\\
CGCG 038-051 & 10 55 39.2 & +02 23 45 & dIrr & 19.0 & 1021 & -15.0 & 0.57 $\times$ 0.26 & ---\\
CGCG 041-023 & 12 01 44.3 & +05 49 17 & SB ? & 23.3 & 1350 & -16.7 & 0.72 $\times$ 0.53 & VV 462\\
SBS 1222+614 & 12 25 05.4 & +61 09 11 & dIrr & 11.4 &  706 & -14.6 & 0.51 $\times$ 0.43 & ---\\
\hline 
\end{tabular}\\
\label{tab:01}
\end{table*}     

In this paper, we present slit-based optical spectroscopic observations of spatially-resolved star-forming regions in four dwarf galaxies. These galaxies are taken from the galaxy catalogue made by \citet{2008A&A...485..657B}, in which these galaxies are classified as Wolf-Rayet (WR) galaxies based on detection of broad emission line features in the optical spectrum from the Sloan Digital Sky Survey (SDSS) data release 6 (DR6). The WR galaxies are a subset of star-bursting \HII emission-line galaxies, which show high ionization emission line of \HeII $\lambda$4686 in their optical spectra along with two broad features around \HeII $\lambda$4686 and C{\small{IV}} $\lambda$5808 emission lines known as the blue bump and the red bump respectively \citep{1976MNRAS.177...91A,1982ApJ...261...64O,1991ApJ...377..115C}. The WR phase in a galaxy is a strong indicator of an ongoing young starburst ($\textless$ 10 Myr) activity, as the most massive O/B-type stars come in to the WR phase after 2 to 5 Myr from their birth before they end this phase through supernovae explosions in a very short time of $\textless$ 0.5 Myr \citep{2005A&A...429..581M}. While the SDSS optical spectra for the selected WR galaxies are already available, we re-observed these galaxies due to the following reasons: (a) the SDSS spectrum has the wavelength coverage of $3800-9200$~\AA, which missed the emission line [O{\small{II}}] $\lambda$3727 for low-z galaxies, required for obtaining a direct estimate for the oxygen abundance, (b) galaxies in our sample contain multiple star-forming regions, and the SDSS spectra were obtained at a single location often coinciding with the brightest \HII region. The general properties of the sample galaxies are provided in Table~\ref{tab:01}. These WR dwarf galaxies were previously studied by \citet{2016MNRAS.462...92J} using the deep H$\alpha$ and SDSS r-band imaging. We present here physical and chemical properties of spatially-resolved star-forming regions in these galaxies.

\section{Observations and data reduction}
\begin{figure*}
\centering
\includegraphics[width=12.0cm,height=11.0cm]{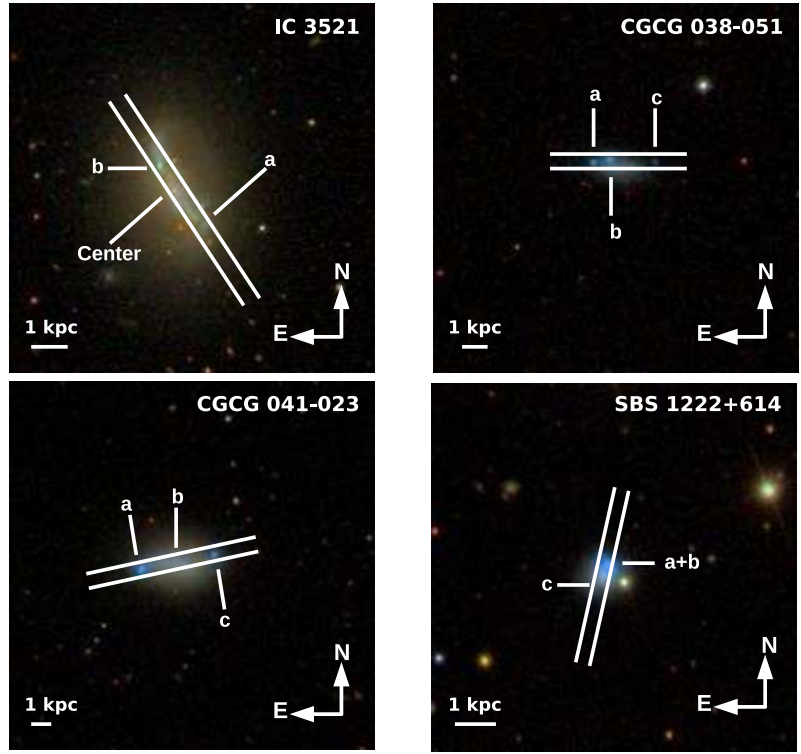}
\caption{The colour composite image made using \textit{g}, \textit{r} and \textit{i}-band images taken from the SDSS survey. The slit positions in the HCT observations are overlaid on the images. The spatially-resolved \HII regions (blue regions) are also labeled over which the spectra are extracted.}
\label{fig:01}
\end{figure*}

\begin{figure*}
\centering
\includegraphics[width=6.0cm,height=8.0cm,angle=270]{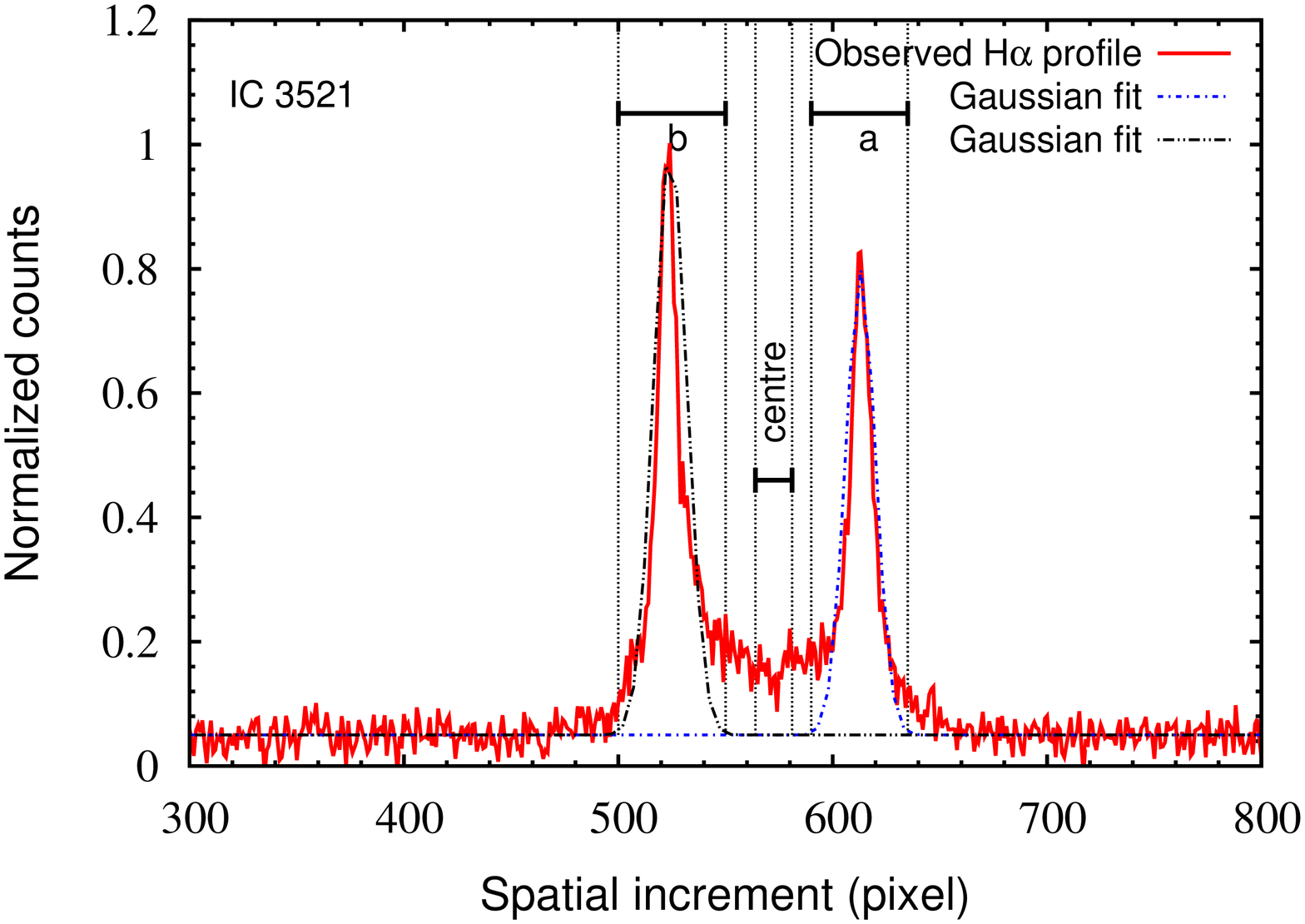}
\includegraphics[width=6.0cm,height=8.0cm,angle=270]{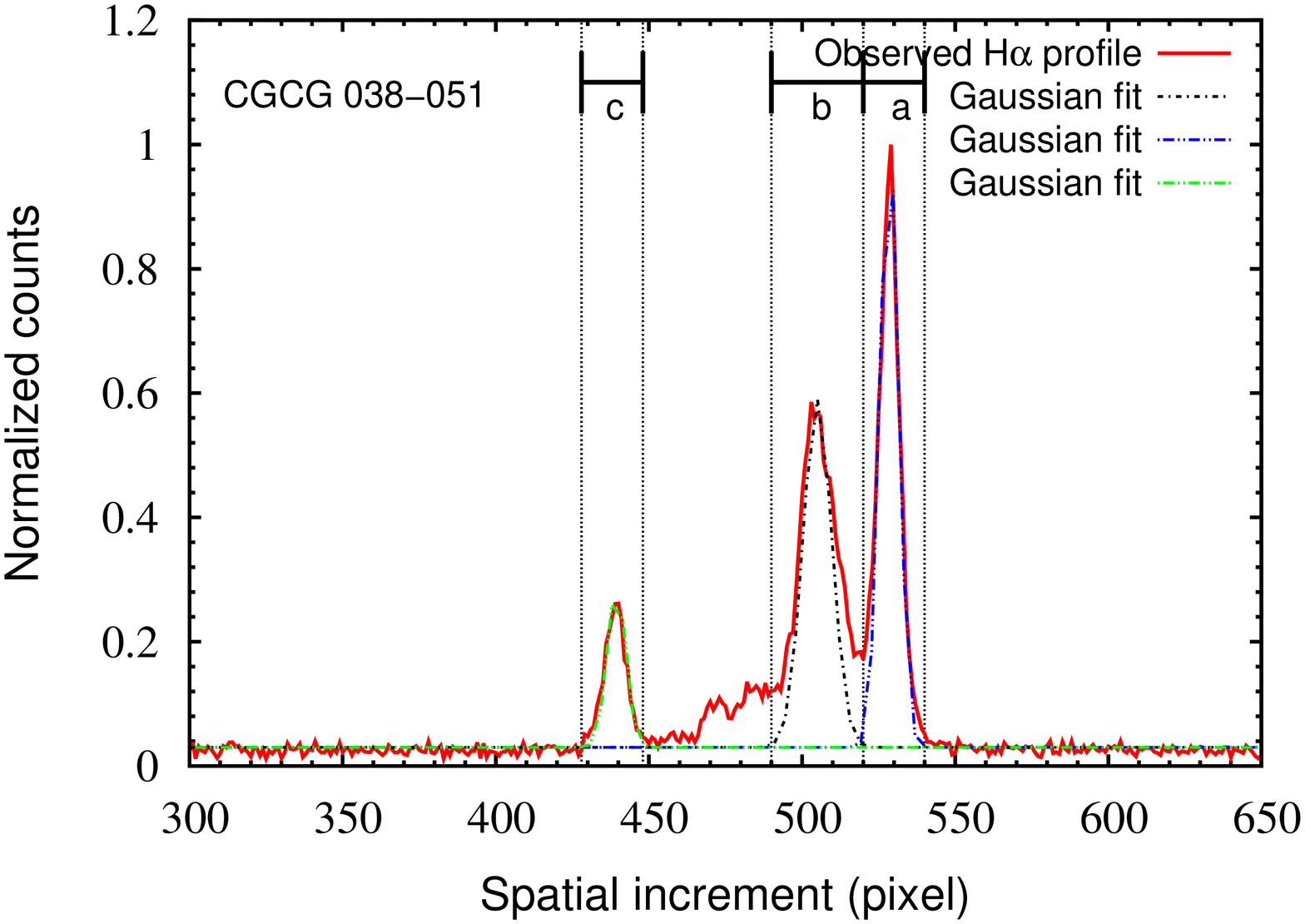}
\includegraphics[width=6.0cm,height=8.0cm,angle=270]{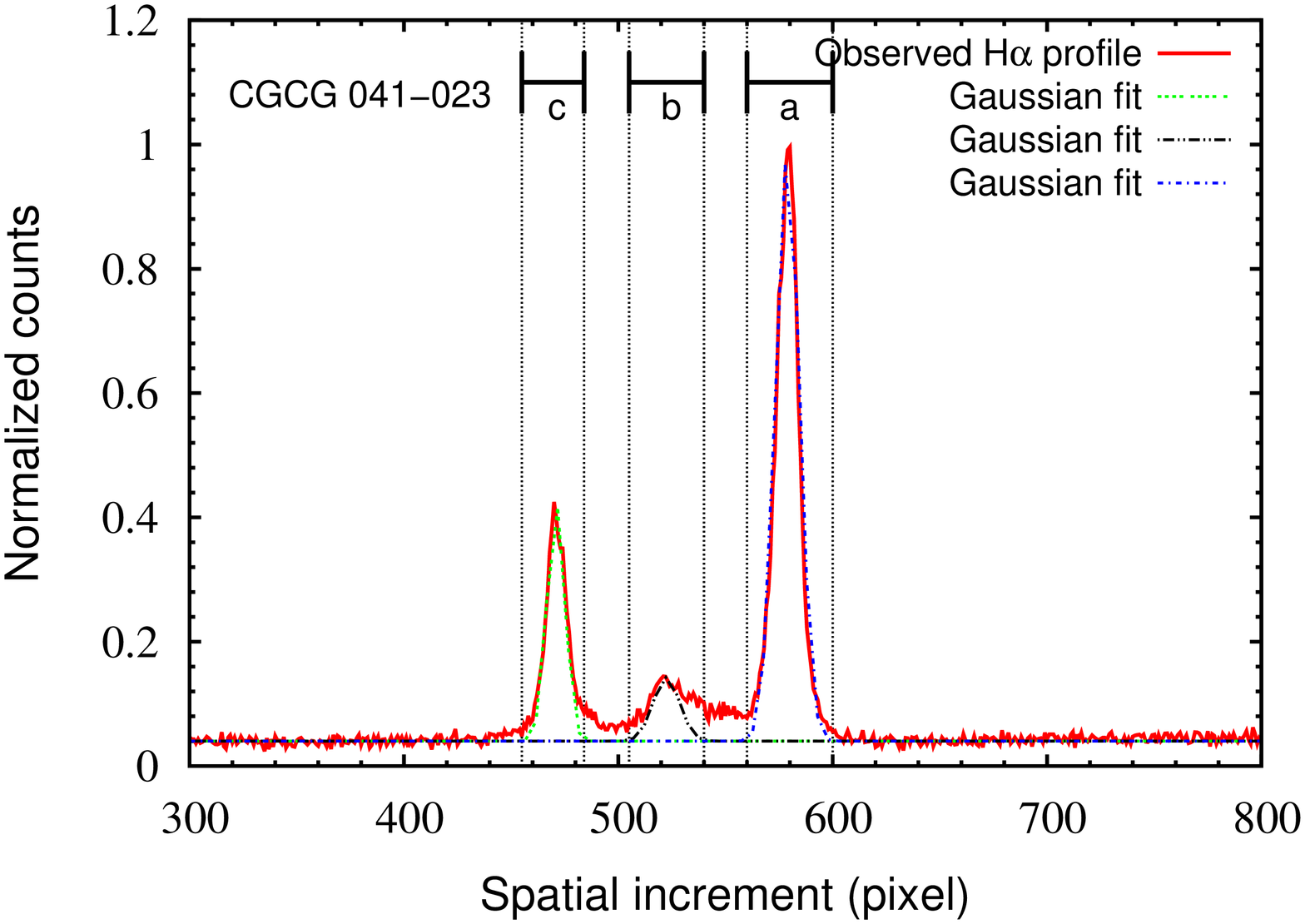}
\includegraphics[width=6.0cm,height=8.0cm,angle=270]{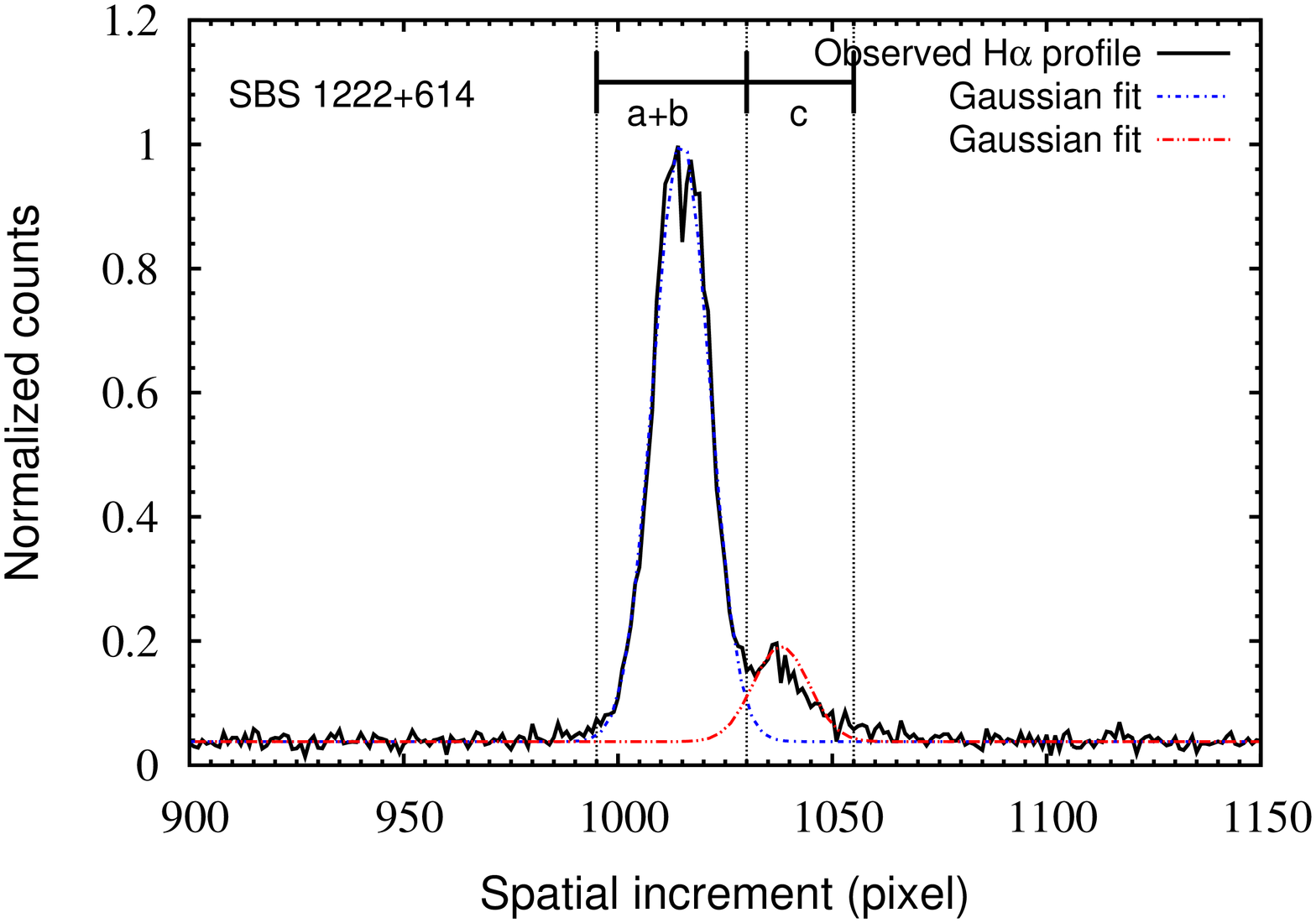}
\caption{The H$\alpha$ profiles of the spatially-resolved \HII regions in our sample of WR galaxies. The size of aperture for each region is also shown for which the spectra are extracted.}
\label{fig:02}
\end{figure*}

The Faint Object Spectrograph and Camera (FOSC) mounted on the 2-m Himalayan Chandra Telescope (HCT) of the Indian Astronomical Observatory (IAO), Hanle, India was used to carry out optical spectroscopic observations. The HCT FOSC is equipped with a 2k $\times$ 4k SITe CCD chip, which uses the central 2k $\times$ 2k region with a plate scale of 0.296$\arcsec$ pixel$^{-1}$ for spectroscopic observations. The gain and readout noise of the CCD camera are 1.22 $e^{-}$ ADU$^{-1}$ and 4.87 $e^{-}$ respectively. The spectroscopic observations of the selected dwarf WR galaxies in our sample were obtained with a slit of aperture 1.92$\arcsec$~$\times$ 11$\arcmin$ and a grism providing a spectral resolution of $\sim$1330. The spectrum covers the wavelength range from $\sim$3500 \AA~to $\sim$7500 \AA~with a dispersion of $\sim$1.5 \AA~pixel$^{-1}$ and an effective spectral resolution of $\sim$11 \AA. The seeing FWHM (full width at half-maximum) was in the range of 1.5$\arcsec$ -- 2.2$\arcsec$ with an average value of $\sim$ 2$\arcsec$. In our observations, the slit position was located in such a way that it covered multiple \HII regions in a galaxy. The slit orientations for each galaxy are shown in Fig.~\ref{fig:01}. The Fe-Ar lamp exposures  were used for the wavelength calibration of the spectrum. The absolute flux calibration was achieved by observing spectrophotometric standard star Feige 34 selected from \citet{1990AJ.....99.1621O}. The observational details for the target sources in our sample are provided in Table~\ref{tab:02}.

\begin{table}
\centering
\caption{Summary of the optical spectroscopic observations.}
\vspace {0.3cm}
\begin{tabular}{cccc} \hline 
Galaxy name &	Date & Exposure time & Airmass\\
\hline
	         &    &	[min] &\\ \hline
IC 3521      & 2016 Dec 02   & 20    & 1.3\\
CGCG 038-051 & 2016 Dec 02   & 60    & 1.4\\
CGCG 041-023 & 2016 Dec 01   & 60    & 2.1\\
SBS 1222+614 & 2013 May 12   & 30    & 1.4\\
\hline 
\end{tabular} \\
\label{tab:02}
\end{table}

The spectroscopic data reduction was performed using the standard procedures in the {\small{IRAF}} (Image Reduction and Analysis Facility). Bias-subtraction and flat-fielding were applied on each frame. Cosmic ray removal was done using the Laplacian kernel detection algorithm \citep{2001PASP..113.1420V}. Extraction of one-dimenional spectra based on optimal extraction algorithm by \citet{1986PASP...98..609H} was carried out. This algorithm provides the optimal signal-to-noise ratio (SNR). The aperture adopted by the extraction algorithm is subjected to non-uniform pixel weights with lower weights to pixels which are far from the central peak of the spatial profile and receive less light from the target source. The fluxes for the emission lines were measured with the {\small{SPLOT}} task of the {\small{IRAF}} by directly summing the flux under the line. This task takes care of the proper subtraction of underlying local continuum flux. Similarly, the line equivalent widths (EWs) were measured in standard way by dividing emission line flux with its corresponding underlying local continuum flux. The errors in the line flux measurements were estimated using $rms$ $\times$ $\sqrt{2 \times N}$, where $N$ is the number of pixels covered in the Gaussian profile of the emission line. The $rms$ was estimated from the line-free region (continuum) on both sides of the emission line. 
 
\section{Results and analysis}

\begin{figure*}
\centering
\includegraphics[width=5.5cm,height=17.0cm,angle=270]{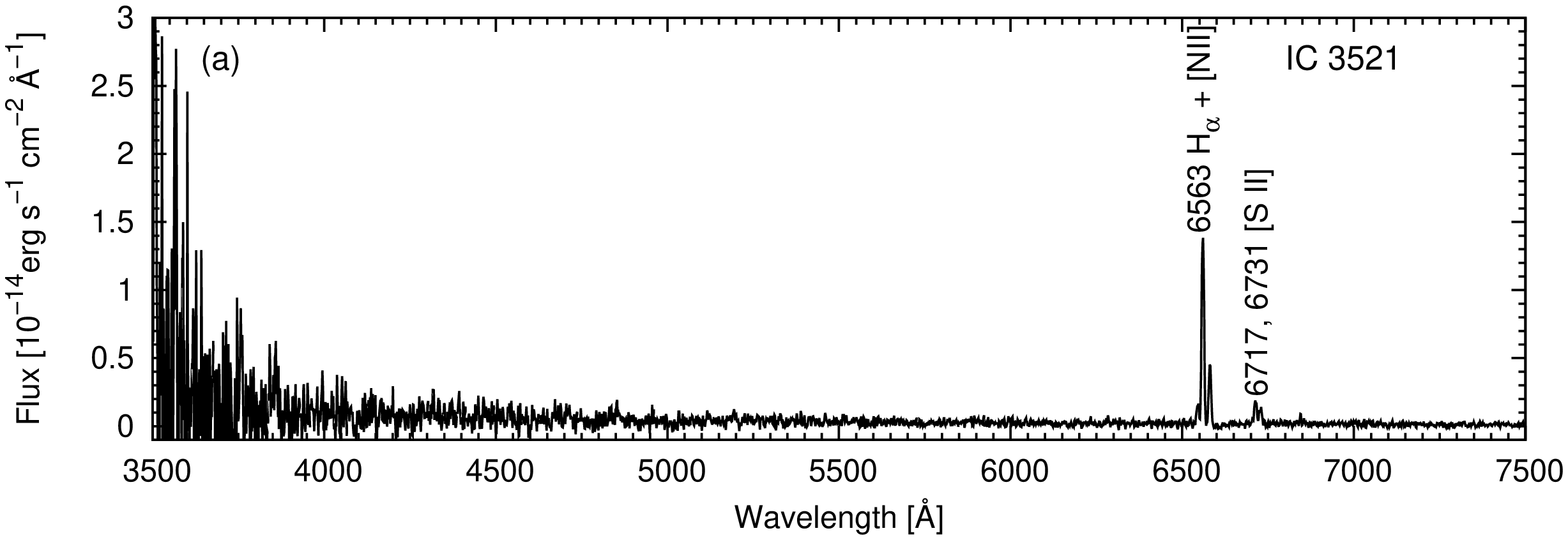}
\includegraphics[width=5.5cm,height=17.0cm,angle=270]{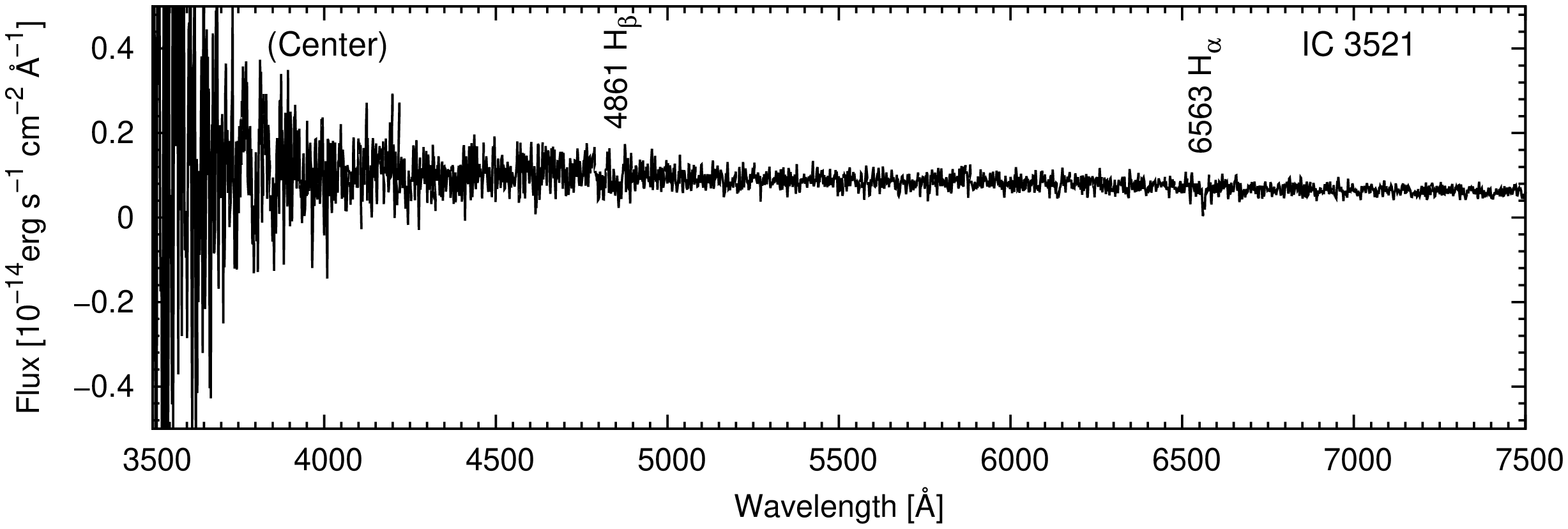}
\includegraphics[width=5.5cm,height=17.0cm,angle=270]{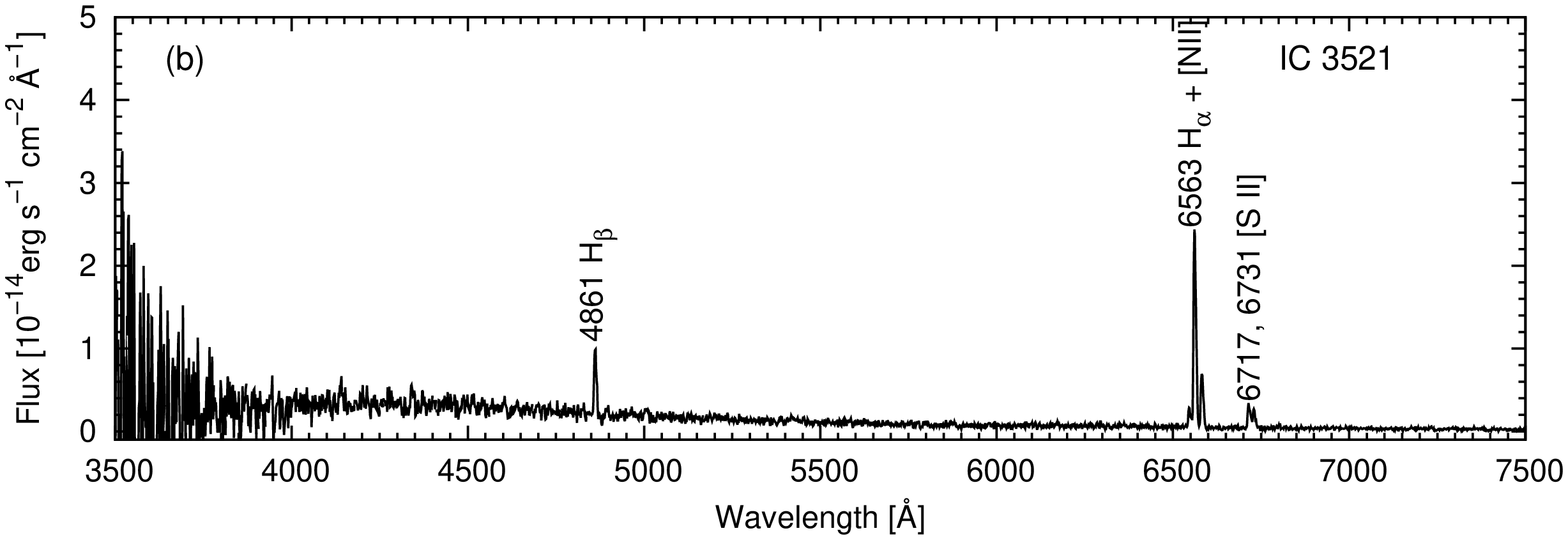}
\caption{The rest-frame extinction-corrected optical spectra of the spatially-resolved \HII regions in dwarf WR galaxy IC 3521, observed with the 2-m HCT.}
\label{fig:03}
\end{figure*}

\begin{figure*}
\centering
\includegraphics[width=5.5cm,height=17.0cm,angle=270]{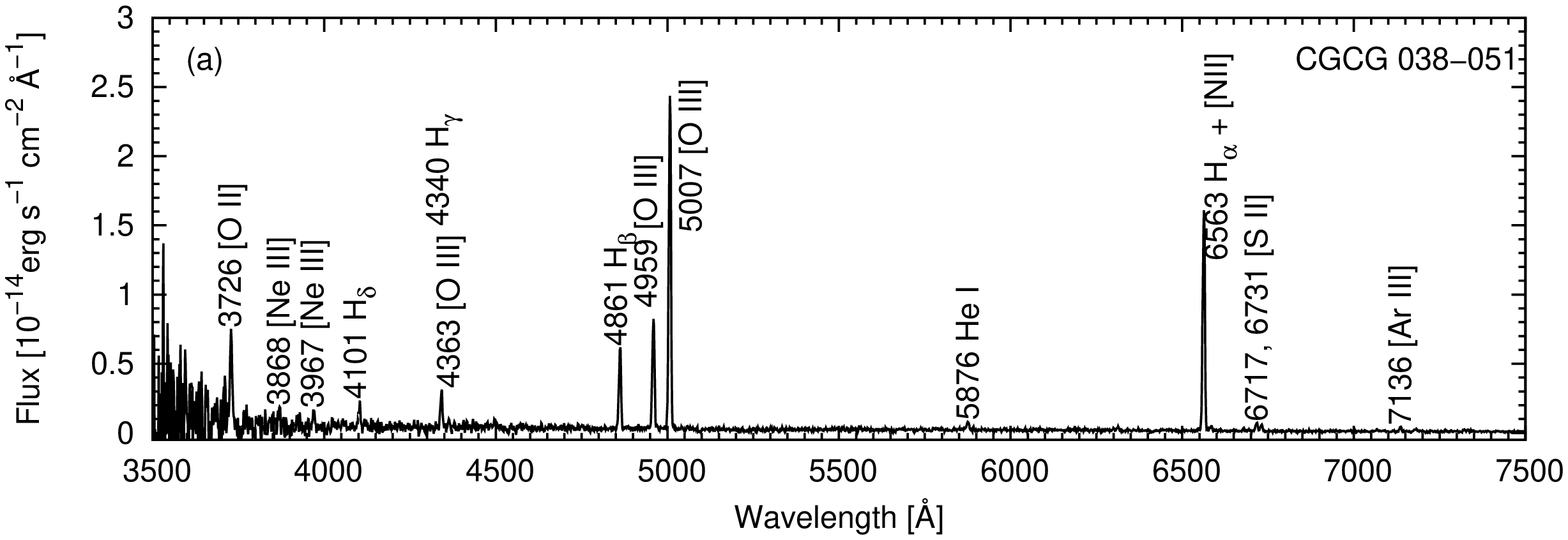}
\includegraphics[width=5.5cm,height=17.0cm,angle=270]{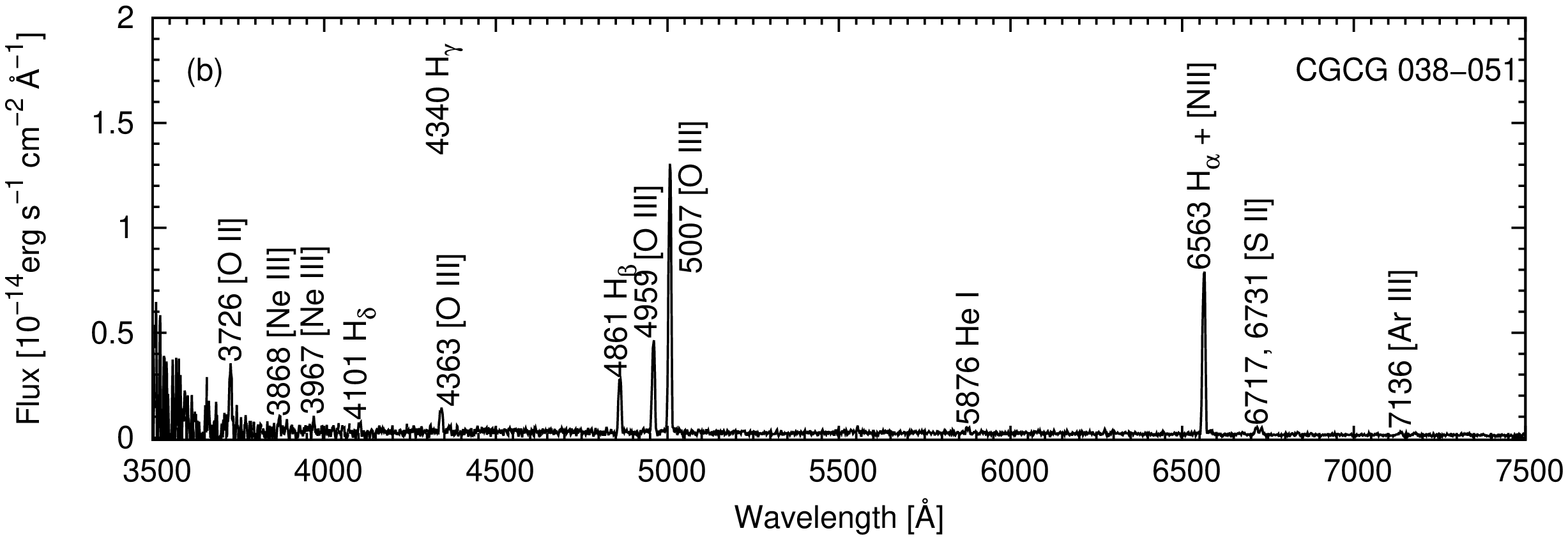}
\includegraphics[width=5.5cm,height=17.0cm,angle=270]{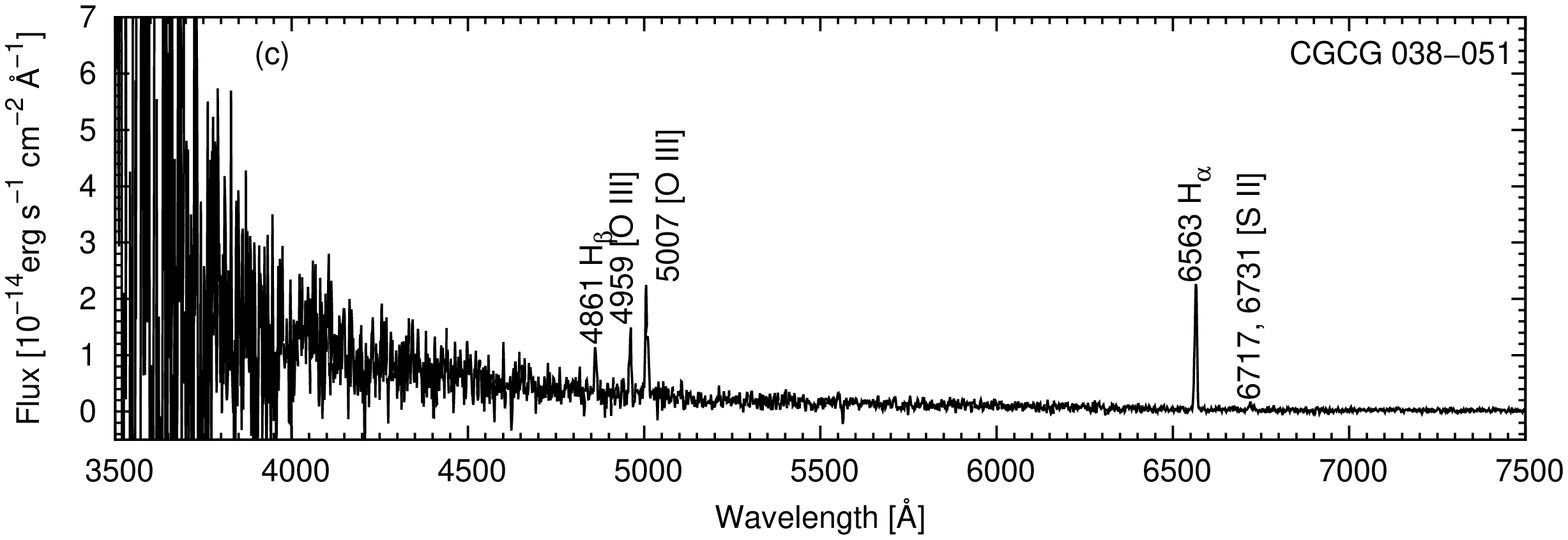}
\caption{The rest-frame extinction-corrected optical spectra of the spatially-resolved \HII regions in dwarf WR galaxy CGCG 038-051, observed with the 2-m HCT.}
\label{fig:04}
\end{figure*}

\begin{figure*}
\includegraphics[width=5.5cm,height=17.0cm,angle=270]{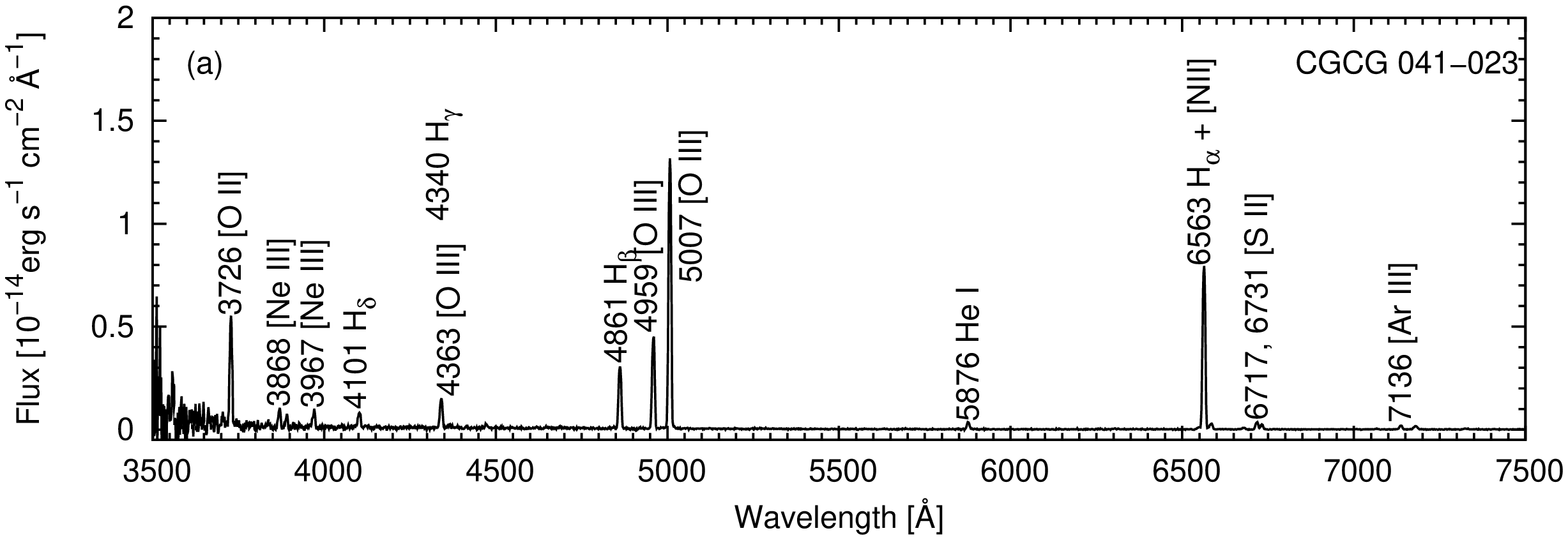}
\includegraphics[width=5.5cm,height=17.0cm,angle=270]{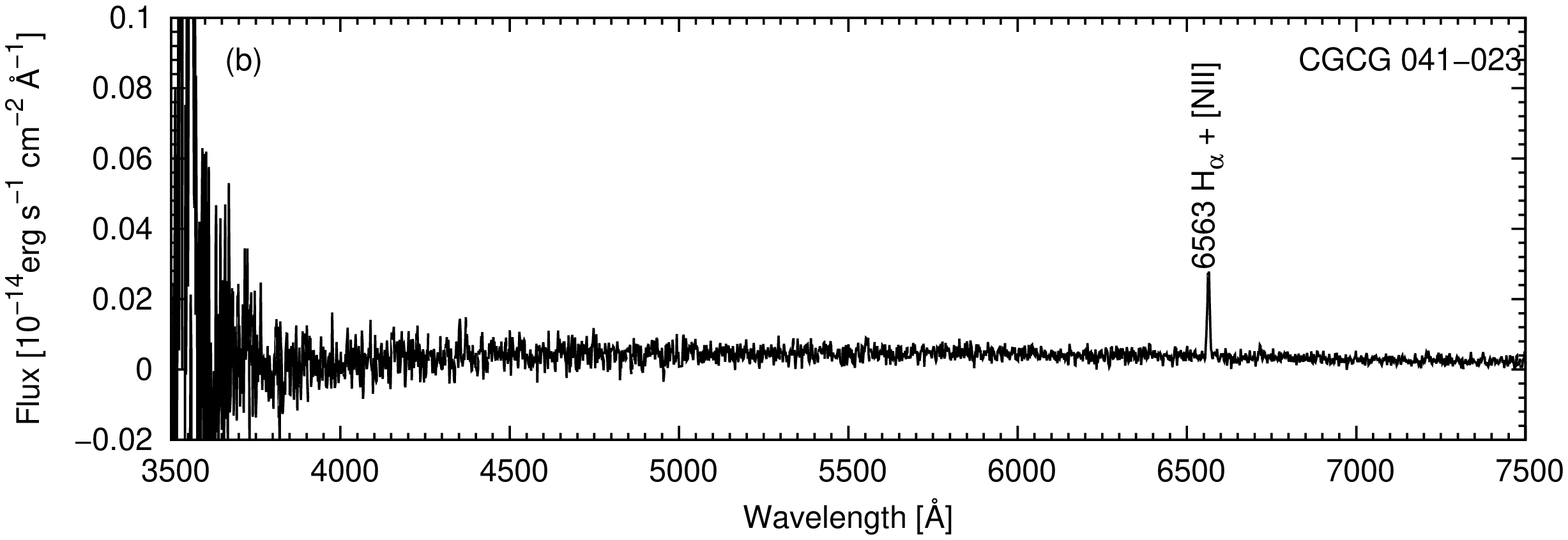}
\includegraphics[width=5.5cm,height=17.0cm,angle=270]{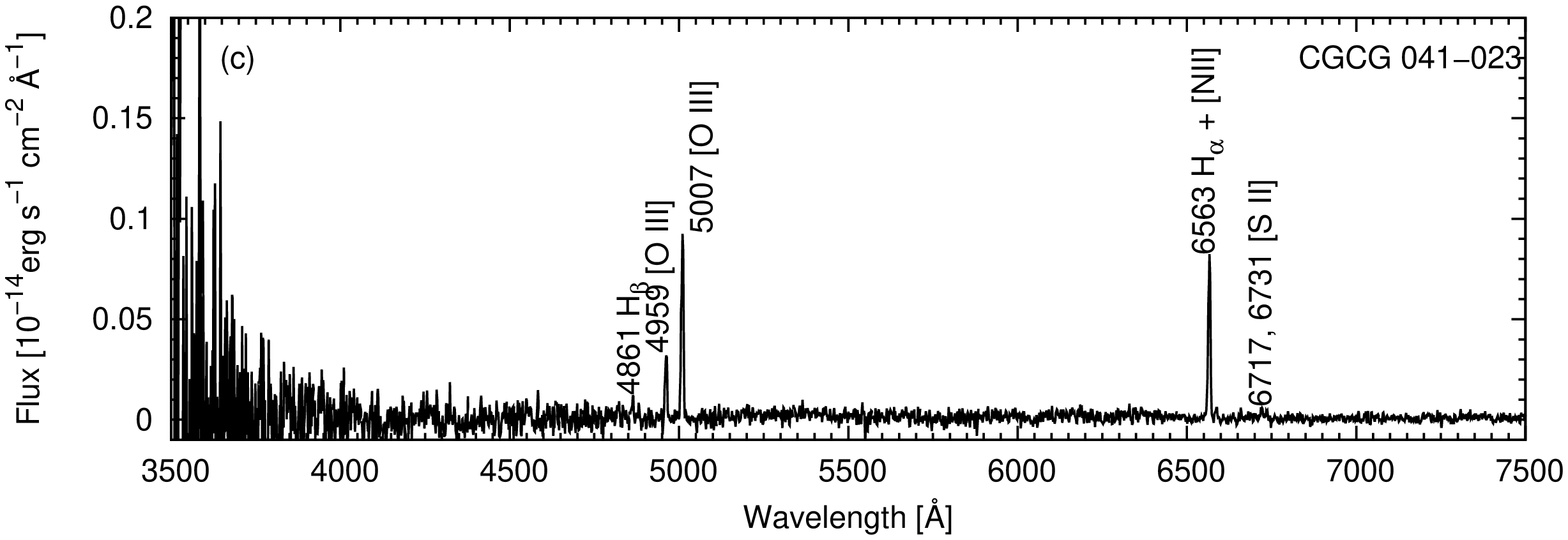}
\caption{The rest-frame extinction-corrected optical spectra of the spatially-resolved \HII regions in dwarf WR galaxy CGCG 041-023, observed with the 2-m HCT.}
\label{fig:05}
\end{figure*}

\begin{figure*}
\centering
\includegraphics[width=5.5cm,height=17.0cm,angle=270]{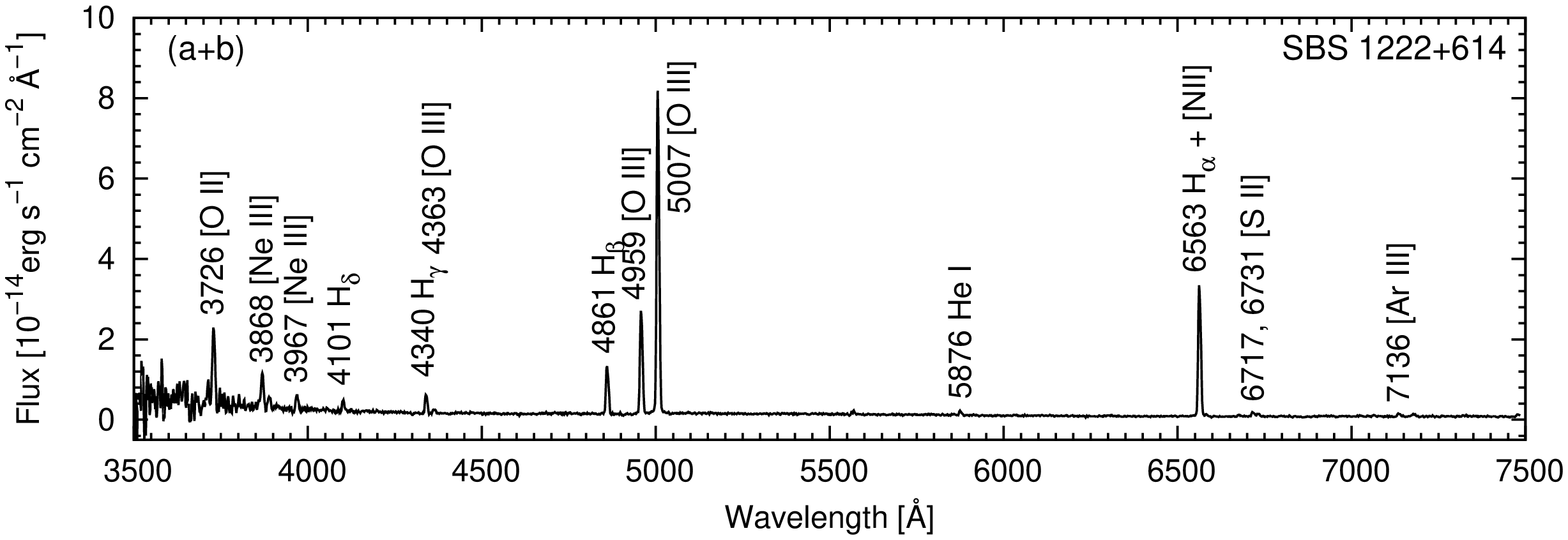}
\includegraphics[width=5.5cm,height=17.0cm,angle=270]{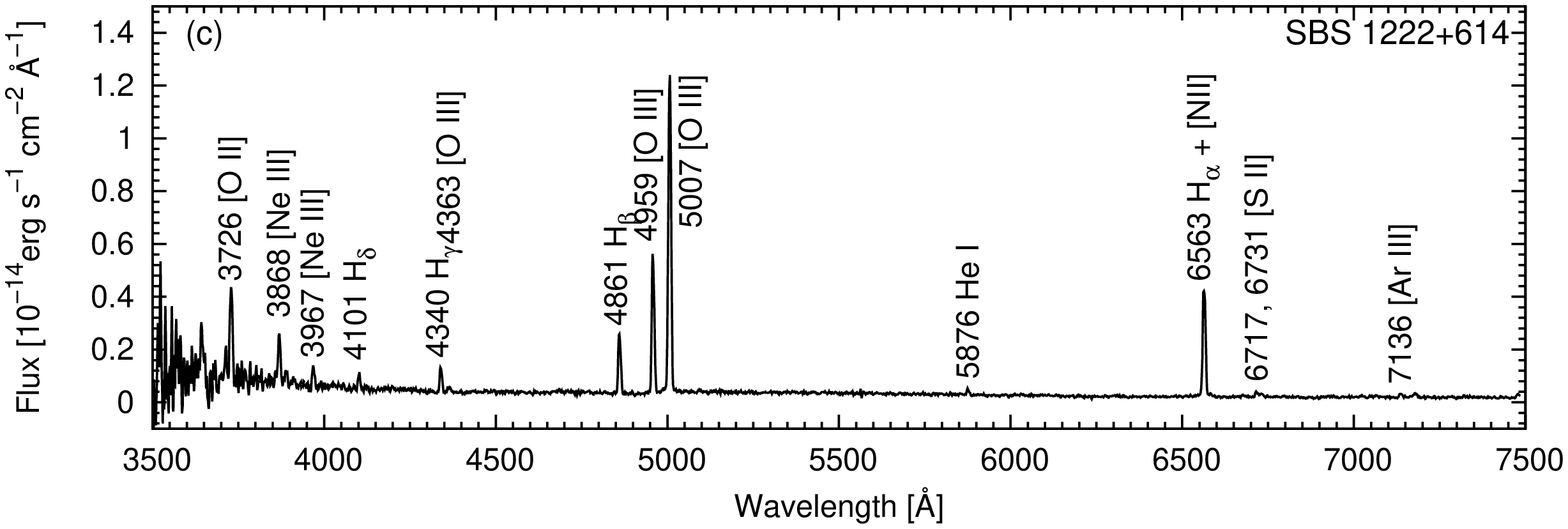}
\caption{The rest-frame extinction-corrected optical spectra of the spatially-resolved \HII regions in dwarf WR galaxy SBS 1222+614, observed with the 2-m HCT.}
\label{fig:06}
\end{figure*}

The star-forming regions in dwarf WR galaxies studied here were previously identified from the continuum subtracted H${\alpha}$ image taken with the 1.3-m Devasthal Fast Optical Telescope \citep[DFOT;][]{2016MNRAS.462...92J}. These \HII regions appear blue in color composite images made using the SDSS $g$, $r$ and $i$-band images as shown in Fig.~\ref{fig:01}. The H$\alpha$ emission profiles and the size of the apertures over which the spectra corresponding to the locations of spatially-resolved \HII regions were extracted, are shown in Fig.~\ref{fig:02}. Several distinct \HII regions having different angular extents can be identified in these images. The optical one-dimensional spectra were extracted over different extraction apertures so as to cover most of the bright blue emitting regions in the galaxies. The sizes of the extraction apertures used here vary from $\sim$ 6$\arcsec$ $\times$ 1.92$\arcsec$ to $\sim$ 14$\arcsec$ $\times$ 1.92$\arcsec$ depending on the extent of the emission regions. The minimum aperture of $\sim$ 6$\arcsec$ corresponds to nearly three times the observed average seeing FWHM. The extinction corrected calibrated spectra in the rest-frame are shown in Fig.~\ref{fig:03}~-~\ref{fig:06}. The calibrated spectra for each \HII region were dereddened for galactic and internal extinction using the reddening law of \citet{1989ApJ...345..245C} with a total-to-selective extinction ratio of R$_{V}$ = 3.1. The spectra were first corrected for the Galactic extinction using the reddening value of E(B -- V)$_{Galactic}$ in the direction of the galaxies estimated from \citet{2011ApJ...737..103S} recalibration of the \citet{1998ApJ...500..525S} infrared-based dust map, as implemented in NASA/IPAC Extragalactic Database (NED). Thereafter, the Galactic extinction corrected spectra were corrected for internal extinction using the flux ratio of f$_{H\alpha}$/f$_{H\beta}$ lines by assuming the expected theoretical value as 2.86 and the Case-B recombination \citep{1989SvA....33..694O,2002A&A...389..845K} with an electron temperature of $\sim$ 10$^{4}$ K and electron density of 100 cm$^{-3}$ . The estimated values of the Galactic reddening E(B -- V)$_{Galactic}$ and internal reddening E(B -- V)$_{internal}$ are provided in Table~\ref{tab:03}. In some cases, the flux ratio of f$_{H\alpha}$/f$_{H\beta}$ lines was found less than the expected theoretical value of 2.86. A low value of f$_{H\alpha}$/f$_{H\beta}$ is often associated with intrinsically low reddening and hence we assumed a E(B -- V)$_{internal}$ as zero for such cases. Previously, lower than theoretical value of f$_{H\alpha}$/f$_{H\beta}$ has been reported in several galaxies \citep{2009A&A...508..615L,2009MNRAS.396...97R,2013MNRAS.433.2764G,2018MNRAS.473.4566P}. Such low value is usually believed as resulted from variations in physical conditions of ionized gas such as high electron temperature or low electron density in the emission region for which the theoretical ratio f$_{H\alpha}$/f$_{H\beta}$ may be less than 2.86 \citep[e.g.,][]{1980IzKry..62...54G,2009A&A...508..615L}. A low value may also result due to error in the line flux calibration and measurement \citep{2006MNRAS.372..961K}. The spectrum from the central region of IC 3521 shows weak H$\alpha$ line in absorption and no emission line. Therefore, the central region of IC 3521 was not included in further analysis. The spectra for the knot \#a in IC 3521 and knots \#b and \#c in CGCG 041-023  have no or weak detection of the H$\beta$ line although the H$\alpha$ and [N{\small{II}}] $\lambda$6584 lines were clearly detected. Therefore, these spectra could not be corrected for internal extinction. 

The observations of CGCG 041-023 were performed at relatively high airmass of 2.1, where slit-light losses due to differential atmospheric refraction become significant. The slit-light losses are expected to be higher towards the bluer wavelengths. The position angle ($\sim$ 70$^{o}$) of the slit orientation in this case was close to the parallactic angle ($\sim$ 60$^{o}$). We also found that after applying the Galactic extinction correction, the flux ratio of f$_{H\alpha}$/f$_{H\beta}$ for the knot \#a in CGCG 041-023 is $\sim$ 2.71 which is slightly less than the expected theoretical value of 2.86. If slit-light losses near the H$\beta$ line were significant, the apparent f$_{H\alpha}$/f$_{H\beta}$ ratio is supposed to increase from the theoretical value. This suggests that the light losses due to differential atmospheric refraction are not significant.
 
The prominent emission lines were identified and marked in the spectra. These lines include the Balmer lines of Hydrogen H${\delta}$, H${\gamma}$, H${\beta}$, H${\alpha}$, \HeII $\lambda$4686 and numerous forbidden emission lines such as [O{\small{II}}] $\lambda$3726, [O{\small{III}}] $\lambda$4363 and [O{\small{III}}] $\lambda$4959, 5007, [N{\small{II}}] $\lambda$6584, [S{\small{II}}] $\lambda$6717, 6731 and some other emission lines such as [Ne{\small{III}}] $\lambda$3868, 3967 and [Ar{\small{III}}] $\lambda$7136. The flux values obtained for the lines along with EWs of the H${\alpha}$, H${\beta}$ and [O{\small{III}}] $\lambda$5007 lines for the \HII regions in the galaxies are given in Table~\ref{tab:03}. It is known that more than $\sim$ 90 per cent contribution to the emission from the \HII regions in BCDs is due to ongoing young burst of star formation \citep{1996A&AS..120..207P,1999AGM....15.P102N,2001ApJS..136..393C,2009A&A...501...75A}. Normally, the Balmer absorption EWs are found $\textless$ 3 \AA~\citep{1999ApJS..125..489G} in star-forming galaxies, which is insignificant compared to the uncertainty in our estimates for the emission line EWs. Therefore, the Balmer emission line EWs are not corrected here for a relatively weak absorption EW due to old stellar population underlying in the emission region. 

\begin{landscape}
\begin{table}
\centering
\caption{The dereddened optical emission line flux, equivalent width, starburst age and other parameters for the spatially-resolved star-forming regions in dwarf WR galaxies.}
\begin{tabular}{ccccccc}
\hline 
Galaxy & & IC 3521 & IC 3521 & CGCG 038-051 & CGCG 038-051 & CGCG 038-051 \\
       & & (knot \#a) & (knot \#b) & (knot \#a) & (knot \#b) & (knot \#c)\\
\hline
Line  & Wavelength & Flux & Flux & Flux  & Flux & Flux\\
      &  [\AA]  & [10$^{-14}$ erg s$^{-1}$ cm$^{-2}$] & [10$^{-14}$ erg s$^{-1}$ cm$^{-2}$] & [10$^{-14}$ erg s$^{-1}$ cm$^{-2}$]  & [10$^{-14}$ erg s$^{-1}$ cm$^{-2}$] & [10$^{-14}$ erg s$^{-1}$ cm$^{-2}$]\\
\hline
O{\sc II}   & 3726  & ---            & ---            & 3.14$\pm$0.35  & 1.78$\pm$0.18  & --- \\
Ne{\sc III} & 3868  & ---            & ---            & 0.46$\pm$0.15  & 0.32$\pm$0.07  & --- \\
Ne{\sc III} & 3967  & ---            & ---            & 0.42$\pm$0.13  & 0.24$\pm$0.06  & --- \\
H$\delta$   & 4101  & ---            & ---            & 0.64$\pm$0.08  & 0.19$\pm$0.06  & --- \\
H$\gamma$   & 4340  & ---            & ---            & 1.16$\pm$0.08  & 0.72$\pm$0.05  & --- \\
O{\sc III}  & 4363  & ---            & ---            & 0.23$\pm$0.03  & 1.39$\pm$0.04  & --- \\
H$\beta$    & 4861  & ---            & 3.58$\pm$0.14  & 2.63$\pm$0.03  & 1.60$\pm$0.03  & 0.63$\pm$0.08 \\
O{\sc III}  & 4959  & ---            & ---            & 3.78$\pm$0.02  & 2.50$\pm$0.03  & 0.75$\pm$0.08 \\
O{\sc III}  & 5007  & ---            & ---            & 11.17$\pm$0.04 & 7.36$\pm$0.03  & 1.59$\pm$0.09\\
He{\sc I}   & 5876  & ---            & ---            & 0.30$\pm$0.02  & 0.18$\pm$0.02  & --- \\
O{\sc I}    & 6300  & ---            & ---            & 0.05$\pm$0.01  & 0.05$\pm$0.01  & --- \\
N{\sc II}   & 6548  & 0.69$\pm$0.04  & 0.86$\pm$0.03  & ---            & ---            & --- \\
H$\alpha$   & 6563  & 6.21$\pm$0.05  & 10.20$\pm$0.04 & 7.50$\pm$0.02  & 4.55$\pm$0.01  & 1.79$\pm$0.02\\
N{\sc II}   & 6583  & 1.92$\pm$0.05  & 2.86$\pm$0.04  & 0.17$\pm$0.01  & 0.12$\pm$0.01  & --- \\
S{\sc II}   & 6717  & 0.81$\pm$0.04  & 1.27$\pm$0.05  & 0.28$\pm$0.01  & 0.24$\pm$0.01  & 0.09$\pm$0.02 \\
S{\sc II}   & 6731  & 0.54$\pm$0.03  & 1.06$\pm$0.04  & 0.20$\pm$0.01  & 0.23$\pm$ 0.01 & 0.03$\pm$0.01 \\  
Ar{\sc III} & 7135  & ---            & ---            & 0.14$\pm$0.01  & 0.10$\pm$0.01  & 0.03$\pm$ 0.01\\
\hline
$E(B-V)_{foreground}$                  && 0.02            & 0.02           & 0.03            & 0.03            &
0.03\\
$E(B-V)_{internal}$                    && 0               & 0.19$\pm$0.04  & 0.15$\pm$0.01   & 0               &  
0.58$\pm$0.09\\
M$_{B}$                                && -12.07$\pm$0.16 & -9.01$\pm$0.16 & -12.20$\pm$0.15 & -14.88$\pm$0.15 &
-10.66$\pm$0.15\\
$u-r$                                  && 0.23$\pm$0.01   & 1.69$\pm$0.14  & 0.40$\pm$0.04   & 0.86$\pm$0.02   &
0.13$\pm$0.11\\ 
Log (M$_{stellar}$) [M$_{\odot}$]      && 6.95            & 7.21           & 6.12            & 7.89                          & 6.67 \\
- EW (H$\alpha$) [\AA]                 && 525$\pm$109    & 369$\pm$27      & 767$\pm$60      & 491$\pm$34      & 485$\pm$82 \\
- EW (H$\beta$) [\AA]                  && ---            & 30$\pm$2        & 139$\pm$10      & 117$\pm$13      & 18$\pm$3 \\
- EW ([O{\sc III}]$\lambda$5007) [\AA] && ---            & ---             & 643$\pm$51    & 575$\pm$45    & 59$\pm$9 \\
Log [O III] $\lambda$5007/H$\beta$     && ---            & ---             & 0.63$\pm$0.01  & 0.66$\pm$0.01  & 0.40$\pm$0.06 \\
Log [N II] $\lambda$6583/H$\alpha$     && -0.51$\pm$0.01 & -0.55$\pm$0.01  & -1.64$\pm$0.03 & -1.58$\pm$0.04 & --- \\
Log [S II] $\lambda$6717, 6731/H$\alpha$     && -0.66$\pm$0.08 & -0.64$\pm$0.01 & -1.19$\pm$0.01 & -1.00$\pm$0.01 & -1.15$\pm$0.08 \\
$^{a}$P-parameter                      && ---            & ---            & 0.82$\pm$0.02  & 0.85$\pm$0.03  & --- \\
$^{b}$R$_{23}$-parameter               && ---            & ---            & 6.92$\pm$0.16  & 7.24$\pm$0.18  & --- \\
$^{c}$R$_{3}$-parameter                && ---            & ---            & 5.67$\pm$0.07  & 6.18$\pm$0.12  & --- \\
Age [Myr]                              && 4.01$\pm$0.48    & 5.31$\pm$0.04    & 6.61$\pm$0.16    & 8.01$\pm$0.26    & 10.74$\pm$0.29 \\
\hline 
\end{tabular}
\begin{flushleft}
$^{a,b,c}$ These parameters are defined in Sect.~\ref{sect:oxyabnd}.
\end{flushleft}
\label{tab:03}
\end{table}
\end{landscape} 

\begin{landscape}
\begin{table}
\centering
\begin{tabular}{cccccccc}
\hline 
Galaxy & & CGCG 041-023 & CGCG 041-023 & CGCG 041-023 & SBS 1222+614 & SBS 1222+614 \\
       & & (knot \#a) & (knot \#b) & (knot \#c) & (knot \#a+b) & (knot \#c)\\
\hline
Line  & Wavelength &  Flux & Flux & Flux & Flux  & Flux\\
      &  [\AA]  & [10$^{-14}$ erg s$^{-1}$ cm$^{-2}$] & [10$^{-14}$ erg s$^{-1}$ cm$^{-2}$] & [10$^{-14}$ erg s$^{-1}$ cm$^{-2}$] & [10$^{-14}$ erg s$^{-1}$ cm$^{-2}$]  & [10$^{-14}$ erg s$^{-1}$ cm$^{-2}$]\\
\hline
O{\sc II}   & 3726 & 4.61$\pm$0.11   & ---            & ---             & 18.2$\pm$0.8  & 3.24$\pm$0.17\\
Ne{\sc III} & 3868 & 0.64$\pm$0.08   & ---            & ---             & 7.3$\pm$0.5   & 1.51$\pm$0.18\\
Ne{\sc III} & 3967 & 0.63$\pm$0.05   & ---            & ---             & 3.6$\pm$0.2   & 0.70$\pm$0.05\\
H$\delta$   & 4101 & 0.67$\pm$0.04   & ---            & ---             & 2.2$\pm$0.2   & 0.42$\pm$0.04\\
H$\gamma$   & 4340 & 1.34$\pm$0.03   & ---            & ---             & 4.3$\pm$0.1   & 0.87$\pm$0.03\\
O{\sc III}  & 4363 & 0.14$\pm$0.02   & ---            & ---             & 1.1$\pm$0.1   & 0.27$\pm$0.03\\
H$\beta$    & 4861 & 2.89$\pm$0.02   & ---            & ---			    & 11.2$\pm$0.1  & 2.10$\pm$0.02\\
O{\sc III}  & 4959 & 4.24$\pm$0.02   & ---            & 0.09$\pm$0.01   & 24.0$\pm$0.1  & 4.82$\pm$0.03\\
O{\sc III}  & 5007 & 12.55$\pm$0.02  & ---            & 0.25$\pm$0.01   & 72.9$\pm$0.1 & 11.90$\pm$0.03\\
He{\sc I}   & 5876 & 0.33$\pm$0.01   & ---            & ---             & 0.9$\pm$0.1   & 0.17$\pm$0.01\\
O{\sc I}    & 6300 & 0.07$\pm$0.01   & ---            & ---             & ---           & ---\\
N{\sc II}   & 6548 & 0.014$\pm$0.006 & ---            & ---             & ---           & ---\\
H$\alpha$   & 6563 & 7.83$\pm$0.01   & 1.17$\pm$0.03  & 0.23$\pm$0.01   & 30.2$\pm$0.1  & 4.24$\pm$0.02\\
N{\sc II}   & 6583 & 0.29$\pm$0.01   & 0.08$\pm$0.03  & 0.014$\pm$0.002 & 0.6$\pm$0.1   & 0.10$\pm$0.01\\
S{\sc II}   & 6717 & 0.34$\pm$0.01   & ---            & 0.013$\pm$0.001 & 1.1$\pm$0.1   & 0.20$\pm$0.01\\
S{\sc II}   & 6731 & 0.22$\pm$0.01   & ---            & 0.008$\pm$0.001 & 0.8$\pm$0.1   & 0.15$\pm$0.01\\  
Ar{\sc III} & 7135 & 0.18$\pm$0.01   & ---            & ---             & 0.8$\pm$0.1   & 0.14$\pm$0.01\\
\hline
$E(B-V)_{foreground}$                  && 0.01            & 0.01            & 0.01            & 0.01            &
0.01\\
$E(B-V)_{internal}$                    && 0               & 0               & 0               & 0               &  
0\\
M$_{B}$                                && -13.76$\pm$0.15 & -16.65$\pm$0.15 & -12.50$\pm$0.15 & -15.33$\pm$0.15 &
---\\
$u-r$                                  && 0.05$\pm$0.01   & 1.49$\pm$0.01   & 0.19$\pm$0.03   & 0.79$\pm$0.01   &
---\\ 
Log (M$_{stellar}$) [M$_{\odot}$]      && 6.61            & 8.95           & 6.19             & 7.71            & --- \\
- EW (H$\alpha$) [\AA]                 && 1043$\pm$29    & 62$\pm$4      & 913$\pm$364      & 394$\pm$8     & 214$\pm$6\\
- EW (H$\beta$) [\AA]                  && 333$\pm$30     & ---            & ---             & 76$\pm$2      & 61$\pm$2\\
- EW ([O{\sc III}]$\lambda$5007) [\AA] && 2746$\pm$204    & ---            & 2314$\pm$1851      & 490$\pm$9     & 297$\pm$6\\
Log [O III] $\lambda$5007/H$\beta$     && 0.64$\pm$0.01   & ---            & ---             & 0.81$\pm$0.01  & 0.75$\pm$0.01\\
Log [N II] $\lambda$6583/H$\alpha$     && -1.43$\pm$0.01  & -1.18$\pm$0.18 & -1.22$\pm$0.06  & -1.70$\pm$0.03 & -1.63$\pm$0.04\\
Log [S II] $\lambda$6717, 6731/H$\alpha$     && -1.15$\pm$0.01 & ---       & -1.04$\pm$0.02  & -1.22$\pm$0.01 & -1.10$\pm$0.02\\
$^{a}$P-parameter                      && 0.79$\pm$0.01  & ---            & ---              & 0.85$\pm$0.01  & 0.83$\pm$0.01\\
$^{b}$R$_{23}$-parameter               && 7.41$\pm$0.06  & ---            & ---              & 10.23$\pm$0.12 & 9.55$\pm$0.12\\
$^{c}$R$_{3}$-parameter                && 5.82$\pm$0.04  & ---            & ---              & 8.65$\pm$0.08  & 7.97$\pm$0.08\\
Age [Myr]                              && 4.00$\pm$0.11   & 8.81$\pm$0.12    & 4.86$\pm$0.35        & 5.59$\pm$0.03 & 6.01$\pm$0.04\\
\hline 
\end{tabular}
\end{table}
\end{landscape}

\subsection{Age of the recent starburst}

\begin{figure}
\centering
\includegraphics[width=6.5cm,height=8.5cm,angle=270]{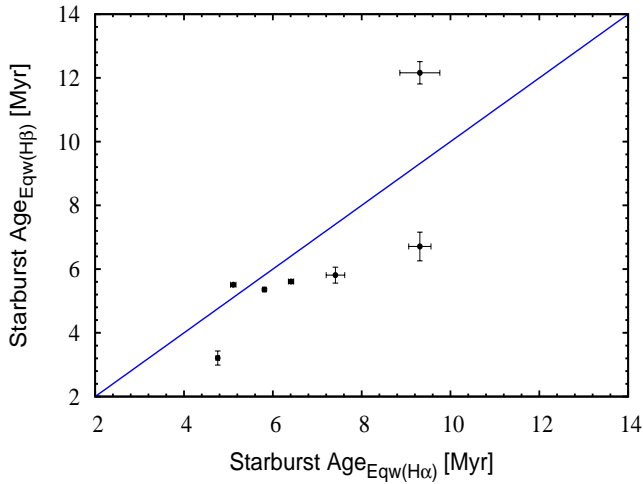}
\caption{A comparison of starburst ages estimated from the H$\alpha$ and H$\beta$ equivalent widths.}
\label{fig:09}
\end{figure}

The age of the most recent star formation can be predicted from the H${\alpha}$ and H${\beta}$ line EWs as the EW decreases with time in a well defined manner \citep{1995ApJ...454L..19L,2000AJ....119.2146J}. We used the Starburst99 model provided by \citet{1999ApJS..123....3L} to estimate the age of the most recent star formation event in our sample of dwarf WR galaxies. The Padova stellar evolutionary model with asymptotic giant branch (AGB) evolution was fitted to obtain EW track of the starburst, assuming the Salpeter initial mass function (IMF) with lower and upper stellar mass limits as 0.1 M$_{\odot}$ and 100 M$_{\odot}$ respectively. This model uses instantaneous star formation scenario. The metallicity input to the model was provided from the estimates made using direct T$_{e}$-method. For the \HII regions where direct estimates of metallicities were not available, the metallicities estimated from empirical calibration based on photoionization models given by \citet{2004MNRAS.348L..59P} were used. The details of the metallicity estimations are presented in Sect.~\ref{sect:oxyabnd}. The EW tracks for the H$\alpha$ and H$\beta$ lines for all spatially-resolved \HII regions were obtained. The age of the star formation was then estimated by comparing the observed EWs of the H$\alpha$ and H$\beta$ lines to that obtained from the model track. A comparison of the star formation age estimated from the H$\alpha$ and the H$\beta$ EWs is shown in Fig.~\ref{fig:09}. This figure suggests that the estimated ages using the observed EWs of the H$\alpha$ and H$\beta$ lines are in good agreement with each other within about $\pm$ 2 Myr. We adopted age of the most recent star formation as the mean of the ages estimated from EWs of the H$\alpha$ and H$\beta$ lines. These ages for the star-forming regions are given in Table~\ref{tab:03}. We found that the age of the most recent starburst in our sample of dwarf WR galaxies is younger than $\sim$6 Myr, except for the regions 'b' and 'c' in CGCG 038-051 and for the region 'b' in CGCG 041-023 which have ages as 8$\pm$1, 10.8$\pm$1.1 and 9$\pm$1 Myr old respectively.

\subsection{Detection of the WR features}

\begin{figure}
\centering
\includegraphics[trim=75 0 0 0, clip, width=6.0cm,height=9.0cm,angle=270]{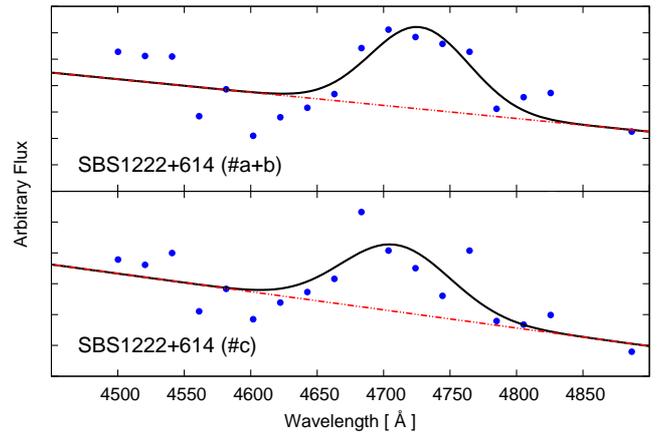}
\caption{The blue WR bump detected in the spatially-resolved star-forming regions of SBS 1222+614. The dashed-dot line represents the continuum fit. The solid line represents a Gaussian fit to the blue bump.}
\label{fig:07}
\end{figure}

\begin{figure}
\centering
\includegraphics[width=3.5cm,height=8.5cm,angle=270]{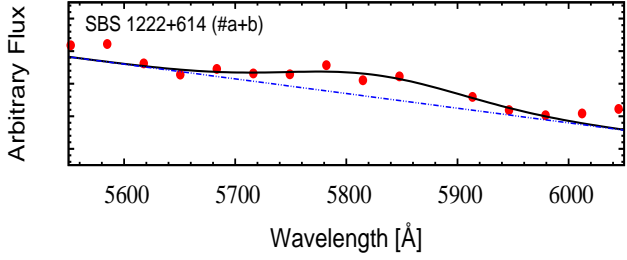}
\caption{The detection of the red WR bump in the spatially-resolved star forming region of SBS 1222+614. The dashed-dot line represents the continuum fit. The solid line represents a Gaussian fit to the red bump.}
\label{fig:08}
\end{figure}

The broad blue WR bump feature around 4686 \AA~was searched in all the star-forming regions in the galaxies. A clear detection of the broad blue WR bump was made only in two star-forming regions hosted by SBS 1222+614 as shown in Fig.~\ref{fig:07}. This detection is in good agreement with \citet{2012MNRAS.421.1043S} who have also reported the WR features in SBS 1222+614. The broad blue bump consists of a blend of C{\small{III}}/C{\small{IV}} $\lambda$4650, 4658, N{\small{III}} $\lambda$4634, 4640, [Ar{\small{IV}}] $\lambda$4711, 4740 and \HeII $\lambda$4686 emission lines. This detection generally indicates a good number ($10^{2}- 10^{5}$) of young WR stars in the galaxy \citep[e.g.,][]{1981A&A...101L...5K,1986A&A...169...71K}. The blue bump appears mainly due to the presence of late-type WN (WNL) and early-type WC (WCE) stars \citep{1998ApJ...497..618S}. The red WR bump around 5808 \AA~is also expected in the WR galaxies. We also possibly identified the red bump feature in the SBS 1222+614 (\# a+b) region as shown in Fig.~\ref{fig:08}. The red WR bump appears mainly due to the presence of emission lines [N{\small{II}}] $\lambda$5795 and He{\small{I}} $\lambda$5875 from the WCE stars. The red bump is rarely detected in WR galaxies as it contains very weak emission line and is expected in high-mettalicity region \citep{2000ApJ...531..776G}. Although we could not detect the WR features in the star-forming regions of other galaxies in our sample, the optical SDSS spectrum for the brightest regions of knot \#b and \#a in IC 3521 and CGCG 0410-023, respectively, shows a detection of WR features \citep{2008A&A...485..657B}. The SDSS spectrum also shows the detection of the broad WR blue bump having a relatively lower strengths in the knot \#b of CGCG 038-051. These detections are missed out in the present study most likely due to the low SNR  in the blue part of the spectra. The starburst ages estimated to be very young ($\leq$ 6 Myr; see Table~\ref{tab:03}) for these star-forming regions including those present in SBS 1222+614 are consistent with the detection of the WR features, which appears only during the very early periods of star formation. Although a few other star-forming regions such as the knot \#a in IC 3521 and CGCG 038-051 and knot \#c in CGCG 041-023 showing a very young starburst of $\leq$ 6 Myr are expected to have the WR features, it is not clear from the present data if these regions also host WR stars or not. Some star-forming regions such as knot \#c and b in CGCG 038-051 and CGCG 041-023, respectively, show the starburst of ages of $\sim$ 10 Myr, indicating that they have probably completed their WR phases. Overall, the detections of the WR features from our own observations and those from the SDSS data from at least one star-forming region in each galaxy in the sample suggest that the sample galaxies are undergoing young massive star formation phase having a significant population of WR stars.

\subsection{Ionization mechanism}

\begin{figure}
\centering
\includegraphics[width=6.5cm,height=8.5cm,angle=270]{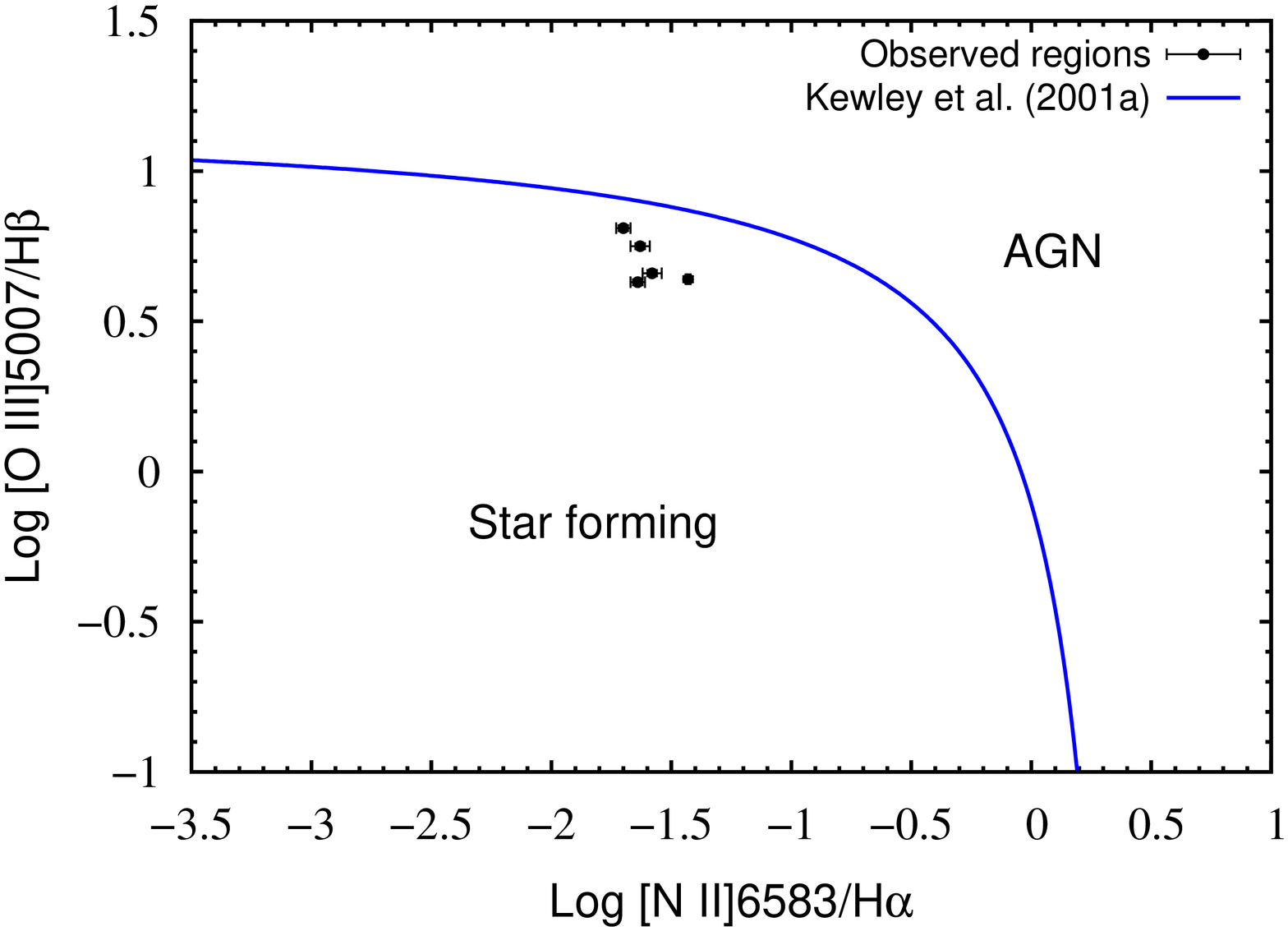}
\includegraphics[width=6.5cm,height=8.5cm,angle=270]{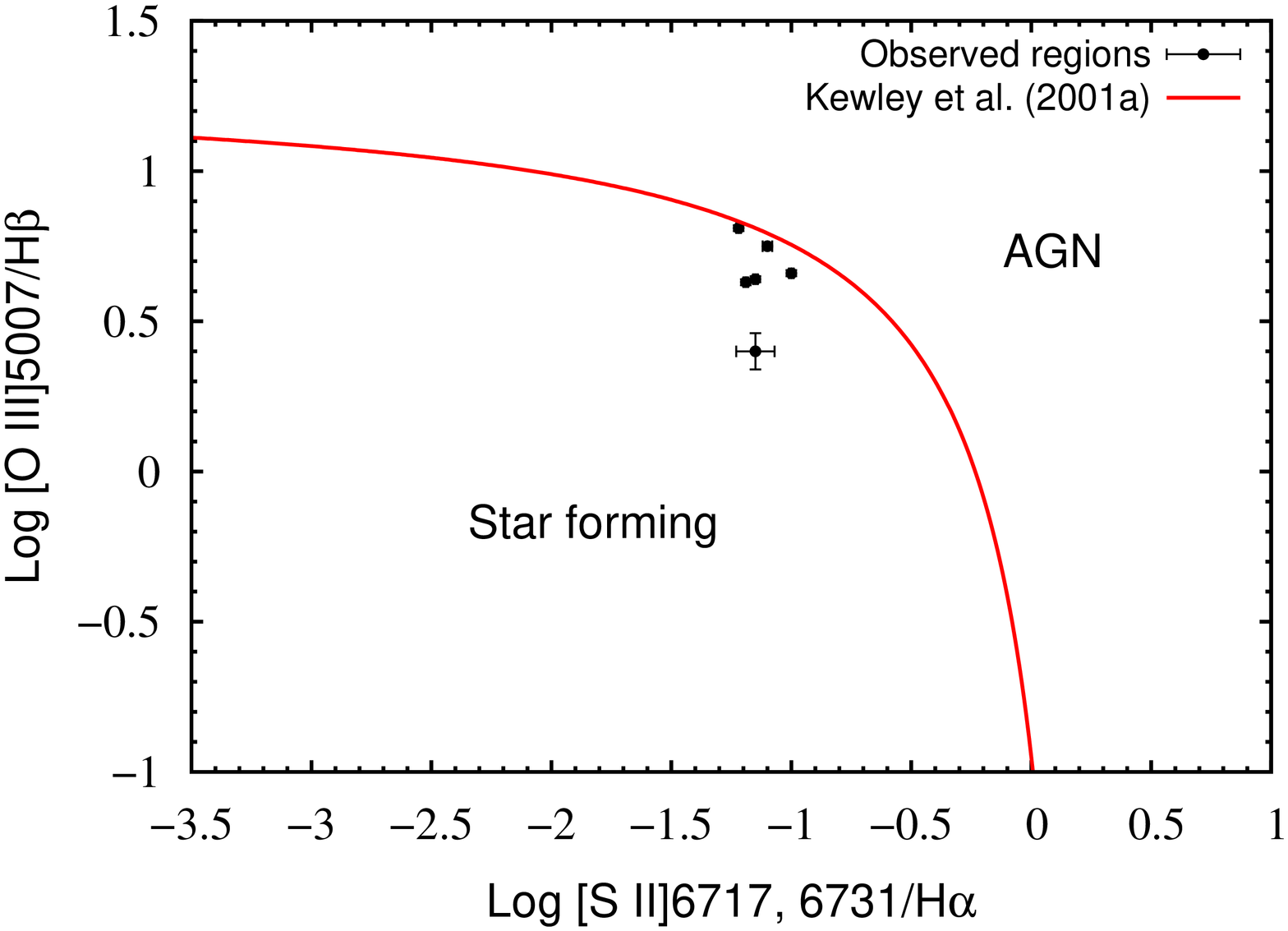}
\caption{A comparison of the observed flux line ratios obtained from the star-forming regions in our sample of dwarf galaxies with the diagnostic diagrams proposed by \citet{2001ApJ...556..121K}.}
\label{fig:10}
\end{figure}

The detection of high ionization emission lines of \HeII $\lambda$4686, [O{\small{III}}] $\lambda$4363, [O{\small{III}}] $\lambda$4959, 5007 and [Ar{\small{III}}] $\lambda$7135 and low ionization emission lines of [O{\small{II}}] $\lambda$3726, [N{\small{II}}] $\lambda$6584 and [S{\small{II}}] $\lambda$6717, 6731 etc, in the galaxy spectrum can be an indication of mix of many excitation sources such as photoionization mainly from massive ionizing stars and shocks generated by WR stars, supernovae explosions and energetic active galactic nuclei (AGN) related mechanisms as discussed by \citet{1991ApJ...373..458G,2004ApJ...607L..21G,2000ApJ...531..776G} and \citet{2010A&A...516A.104L}. Therefore, it is necessary to identify the dominant excitation mechanism in these galaxies. The nature of the dominant ionizing source can be inferred using the quantitative classification scheme proposed by \citet{1981PASP...93....5B} using combinations of line ratios. Subsequently, other similar diagnostic schemes were proposed by \citet{2000ApJ...542..224D} and \citet{2001ApJ...556..121K}. All these diagnostic schemes use the emission line ratios of [O{\small{III}}] $\lambda$5007/H${\beta}$, [N{\small{II}}] $\lambda$6583/H${\alpha}$ and [S{\small{II}}] $\lambda$6717, 6731/H${\alpha}$. The optical emission line ratios for all the spatially-resolved ionized regions in our galaxy sample are listed in Table~\ref{tab:03} and the locations of all the star-forming regions are presented in Fig.~\ref{fig:10} which shows typical plots for [O{\small{III}}] $\lambda$5007/H${\beta}$ versus [N{\small{II}}] $\lambda$6583/H${\alpha}$ (top panel) and [O{\small{III}}] $\lambda$5007/H${\beta}$ versus [S{\small{II}}] $\lambda$6717, 6731/H${\alpha}$ (bottom panel). Here, we used the latest \citet{2001ApJ...556..121K} diagnostic criteria to identify source of excitation mechanism. In these figures, it can be seen that the primary and the dominating source of ionization is photoionization in all the cases. 

\begin{landscape}
\begin{table}
\tiny
\centering
\caption{The physical conditions and chemical abundances of star-forming regions in galaxies.}
\begin{tabular}{ccccccccccc}
\hline 
 Parameters & IC 3521 & IC 3521 & CGCG 038-051 & CGCG 038-051 & CGCG 038-051 & CGCG 041-023 & CGCG 041-023 & CGCG 041-023 & SBS 1222+614 & SBS 1222+614 \\
            & (knot \#a) & (knot \#b) & (knot \#a) & (knot \#b) & (knot \#c) & (knot \#a)   &   (knot \#b) & (knot \#c)   & (knot \#a+b) &  (knot \#c) \\
\hline 

T$_{e}$ [O{\sc III}] (K) & 10000$^{b}$ & 10000$^{b}$ & 16545$\pm$2236  & 16155$\pm$343 & 10000$^{b}$ & 12346$\pm$935    & 10000$^{b}$ & 10000$^{b}$ & 13822$\pm$292  & 14403$\pm$239 \\

T$_{e}$ [O{\sc II}] (K)  & 10000$^{b}$ & 10000$^{b}$ & 14581$\pm$1565  & 14308$\pm$240 & 10000$^{b}$ & 11642$\pm$814    & 10000$^{b}$ & 10000$^{b}$ & 12675$\pm$204  & 13082$\pm$167 \\

n$_{e}$ (cm$^{-3}$)      & 100$^{c}$   & 100$^{c}$   & $\textless$ 100 & 603$\pm$36    & $\textless$ 100 & $\textless$ 100  & 100$^{c}$   & 100$^{c}$   & $\textless$ 100 & $\textless$ 100 \\\\ 

12+log(O$^{++}$/H$^{+}$) & ---           & ---                 & 7.57$\pm$0.11   & 7.63$\pm$0.02 & 8.01$\pm$0.05 & 7.91$\pm$0.08    & ---                 & ---       & 7.95$\pm$0.02 & 7.86$\pm$0.02 \\

12+log(O$^{+}$/H$^{+}$)  & ---           & ---           & 7.08$\pm$0.13   & 7.10$\pm$0.05 & --- & 7.54$\pm$0.10    & ---                 & ---            & 7.42$\pm$0.03 & 7.34$\pm$0.02 \\

12+log(O/H)              & 8.61$\pm$0.01$^{a}$ & 8.59$\pm$0.01$^{a}$ & 7.69$\pm$0.09   & 7.74$\pm$0.02 & 8.01$\pm$0.05 & 8.07$\pm$0.06    & 8.23$\pm$0.10$^{a}$ & 8.20$\pm$0.03$^{a}$       & 8.06$\pm$0.02 & 7.98$\pm$0.01\\\\   

12+log(N$^{+}$/H$^{+}$)  & ---           & 7.18$\pm$0.01       & 5.59$\pm$0.08   & 5.67$\pm$0.04 & --- & 6.01$\pm$0.06    & ---                 & ---       & 5.64$\pm$0.03 & 5.55$\pm$0.04\\

12+log(N/H)              & ---           & 7.66$\pm$0.01       & 6.19$\pm$0.16   & 6.30$\pm$0.06 & --- & 6.53$\pm$0.12    & ---                 & ---       & 6.27$\pm$0.04 & 6.18$\pm$0.05\\

icf(N)                   & ---           & 3$^{d}$             & 4.00$\pm$1.29   & 4.28$\pm$0.45 & --- & 3.30$\pm$0.79    & ---                 & ---          & 4.27$\pm$0.29 & 4.18$\pm$0.27\\

log(N/O)                 & ---           & -0.93$\pm$0.01      & -1.50$\pm$0.18  & -1.44$\pm$0.06 & --- & -1.53$\pm$0.14    & ---                 & ---      & -1.79$\pm$0.05& -1.80$\pm$0.05\\\\

12+log(Ne$^{++}$/H$^{+}$)& ---           & ---            & 6.83$\pm$0.17   & 6.88$\pm$0.07 & --- & 7.34$\pm$0.10    & ---           & ---              & 7.52$\pm$0.03 & 7.50$\pm$0.04\\

12+log(Ne/H)             & ---           & ---                 & 6.88$\pm$0.17   & 6.92$\pm$0.08 & --- & 7.39$\pm$0.11    & ---           & ---                 & 7.56$\pm$0.03 & 7.54$\pm$0.04\\

icf(Ne)                  & ---           & ---              & 1.10$\pm$0.10  & 1.10$\pm$0.02 & --- & 1.13$\pm$0.07      & ---           & ---              & 1.10$\pm$0.02  & 1.10$\pm$0.01\\

log(Ne/O)                & ---           & ---              & -0.81$\pm$0.19  & -0.82$\pm$0.08& --- & -0.67$\pm$0.12     & ---           & ---              & -0.50$\pm$0.04  & -0.43$\pm$0.04\\\\

12+log(S$^{+}$/H$^{+}$)  & ---           & 6.16$\pm$0.02       & 5.27$\pm$0.07   & 5.50$\pm$0.02 & 5.63$\pm$0.09 & 5.48$\pm$0.06    & ---                 & ---       & 5.35$\pm$0.02 & 5.32$\pm$0.02\\

12+log(S/H)              & ---           & 6.64$\pm$0.02       & 5.88$\pm$0.18   & 6.14$\pm$0.06 & 6.23$\pm$0.09 & 6.01$\pm$0.13    & ---                 & ---       & 5.99$\pm$0.04 & 5.95$\pm$0.04\\

icf(S)                   & ---           & 3$^{d}$             & 4.09$\pm$1.51   & 4.40$\pm$0.53 & 4$^{d}$ & 3.33$\pm$0.92    & ---       & ---          & 4.39$\pm$0.34 & 4.29$\pm$0.31\\

log(S/O)                 & ---           & -1.95$\pm$0.02      & -1.80$\pm$0.20  & -1.60$\pm$0.06 & -1.78$\pm$0.10 & -2.06$\pm$0.15   & ---        & ---      & -2.06$\pm$0.04 & -2.03$\pm$0.04\\\\

12+log(Ar$^{++}$/H$^{+}$) & ---      & ---    & 5.26$\pm$0.08   & 5.33$\pm$0.05 & 5.57$\pm$0.15 & 5.57$\pm$0.06    & ---              & ---                 & 5.52$\pm$0.04 & 5.46$\pm$0.03\\

12+log(Ar/H)      & ---           & ---      & 5.25$\pm$0.09   & 5.32$\pm$0.05 & 5.57$\pm$0.15 & 5.53$\pm$0.06    & ---                 & ---                 & 5.52$\pm$0.04 & 5.45$\pm$0.03\\

icf(Ar)      & ---           & ---                 & 0.97$\pm$0.08   & 0.99$\pm$0.03 & 1$^{d}$ & 0.92$\pm$0.05    & ---                 & ---       & 0.99$\pm$0.02 & 0.99$\pm$0.02\\

log(Ar/O)       & ---           & ---    & -2.44$\pm$0.12  & -2.42$\pm$0.05 & -2.44$\pm$0.16 & -2.54$\pm$0.09   & ---        & ---      & -2.54$\pm$0.04 & -2.52$\pm$0.04\\

\hline 
\end{tabular}  
\label{tab:04}
\begin{flushleft}
$^a$ The values of 12+log(O/H) which are estimated using N$_{2}$-method, as oxygen lines are not detected in the spectrum.\\
$^b$ The values of electron temperatures that could not be estimated due to the absence of oxygen [O III]$\lambda$4363 line are assumed to be 10000 K.\\
$^c$ The values of electron densities that could not be estimated due to the absence of doublet sulfur [S II]$\lambda\lambda$6717, 6731 lines are assumed to be 100 cm$^{-3}$.\\
$^d$ The values of ICFs that could not be estimated due to the absence of oxygen [O II]$\lambda$3726 line are assumed to be either equal to an approximate value found in other regions within the same galaxy or an average value found in the literature if it is not available for any region of the galaxy. 
\end{flushleft} 
\end{table}
\end{landscape}

\subsection{Physical conditions of the ionized gas}

The electron temperature (T$_{e}$) and density (n$_{e}$) of the ionized gas in the \HII regions are estimated here. Since the faint auroral [O{\small{III}}] $\lambda$4363 emission line along with the [O{\small{III}}] $\lambda$4959, 5007 emission lines were detected in many cases, the electron temperature could be estimated directly using their relative line intensities. A two-zone approximation was assumed to estimate T$_{e}$ for the ionized regions, where T$_{e}$[O{\small{III}}] and T$_{e}$[O{\small{II}}] were taken as representative temperatures for high and low ionization potential ions respectively. We inferred T$_{e}$[O{\small{III}}] from the diagnostic line ratio of [O{\small{III}}] I($\lambda$4959 + $\lambda$5007)/I($\lambda$4363) by using the five-level program within the {\small{NEBULAR}} task of the {\small{IRAF}} for the emission line nebulae \citep{1995PASP..107..896S}. Once T$_{e}$[O{\small{III}}] was estimated, T$_{e}$[O{\small{II}}] was inferred using the linear relation between T$_{e}$[O{\small{III}}] and T$_{e}$[O{\small{II}}] \citep{1992AJ....103.1330G}. The values of T$_{e}$[O{\small{III}}] and T$_{e}$[O{\small{II}}] estimated for the star-forming regions are given in Table~\ref{tab:04}. These derived temperatures are in good agreement with those measured in other nearby star-forming dwarf galaxies \citep[e.g.,][]{1986MNRAS.223..811C,1994ApJ...420..576M,2004ApJ...616..752L,2008MNRAS.385..543H}. In order to measure the electron density of the ionized gas, the diagnostic line ratio of the doublet [S{\small{II}}] I($\lambda$6717)/I($\lambda$6731) was used. Since these two lines of the same ion are emitted from different levels with nearly same excitation energy, the electron density can be estimated using this line ratio. The computed electron densities are listed in Table~\ref{tab:04}.

\subsection{Oxygen abundance} \label{sect:oxyabnd}

We derived the ionic oxygen abundance through electron temperature sensitive lines such as [O{\small{III}}] $\lambda$4363 and [O{\small{III}}] $\lambda$4959, 5007 using T$_{e}$-method as expressed in \citet{2006A&A...448..955I}. Since this method uses the electron temperature-sensitive emission line of oxygen [O{\small{III}}] $\lambda$4363 \citep{2003ApJ...591..801K} which is often very weak and difficult to detect in galaxies, the estimation of ionic oxygen abundance could be made for 5 out of 10 spatially-resolved \HII regions in the galaxies. The estimated ionic oxygen abundances using T$_{e}$-method are given in Table~\ref{tab:04}. This method assumes two zone approximation: a high-ionization zone represented with the temperature T$_{e}$[O{\small{III}}], responsible for [O{\small{III}}] lines; and a low-ionization zone represented with the temperature T$_{e}$[O{\small{II}}], responsible for [O{\small{II}}] lines. The two-zone approximation model for T$_{e}$ is more realistic interpretation of the temperature structure within the \HII regions and provides more accurate estimates of ionic oxygen abundances \citep{2001A&A...369..594P}. The total oxygen abundance is then determined by performing a simple sum of [O{\small{II}}] and [O{\small{III}}] emission lines as follows \citep[e.g.][]{1992MNRAS.255..325P,2005A&A...437..849S,2010A&A...517A..85L,2016ApJ...822..108H}:

\begin{equation}
\frac{{\rm O}}{{\rm H}} = \frac{{\rm O^{+}}}{{\rm H^{+}}} + \frac{{\rm O^{++}}}{{\rm H^{+}}}
\end{equation}

We have also derived the ionic abundances for other elements using the expression given by \citet{2006A&A...448..955I}. Here, we again used the two-zone scheme for determining the ionic abundances. T$_{e}$[O{\small{III}}] is taken as the representative temperature for the high ionization potential ions such as Ne$^{++}$ and Ar$^{++}$, while T$_{e}$[O{\small{II}}] is taken as the representative temperature for the low ionization potential ions such as N$^{+}$ and S$^{+}$. The ionization correction factors (ICF) of \citet{2006A&A...448..955I} were used to compute total abundances for N, Ne and Ar. In case of S, we followed the correction given in \citet{1969BOTT....5....3P}. Subsequently, the log values of N/O, S/O, Ne/O, and Ar/O ratios were also computed. The resulted chemical abundances for the spatially-resolved star-forming region in our sample of dwarf WR galaxies are presented in Table~\ref{tab:04}.

Oxygen abundances corresponding to the \HII regions are also determined using different methods such as N$_{2}$, O$_{3}$N$_{2}$ and P-methods, which are empirically calibrated based on the photoionization models \citep{2001A&A...369..594P,2002MNRAS.330...69D,2004MNRAS.348L..59P}. These empirical methods allow to estimate oxygen abundances where the detection of [O{\small{III}}] $\lambda$4363 emission line is not made. The oxygen abundances obtained using these different methods are given in Table~\ref{tab:05}. A brief overview of these alternate methods is provided below. 

The N$_{2}$ method is mainly based on the N$_{2}$ $\equiv$ log([N{\small{II}}] $\lambda$6583/H$\alpha$ $\lambda$6563) index. Using this index, \citet{1994ApJ...429..572S} provided a tentative calibration for the oxygen abundance. This calibration was revisited by many other investigators including \citet{1998ApJ...497L...1V}, \citet{2000MNRAS.316..559R} and \citet{2002MNRAS.330...69D}. The calibration was subsequently improved after inclusion of spectroscopic measurements of star forming galaxies covering a wide range in metallicity (7.2 $\leq$ 12 + log(O/H) $\leq$ 9.1). More recently, \citet{2004MNRAS.348L..59P} revisited the calibration after inclusion of new extragalactic \HII regions where the values of oxygen abundances in the high and low metallicity regimes were also determined either by the direct T$_{e}$-method or by the detailed photoionization models. The least square linear fit to the data is given by the relation \citep{2004MNRAS.348L..59P}:

\begin{equation}
12 + {\rm log(O/H)} = 8.90 + 0.57{\rm N_{2}}
\end{equation}

Using the above relation, the estimated oxygen abundances for the \HII regions are given in the third column of Table~\ref{tab:05}.

\citet{1979A&A....78..200A} proposed a relation for estimating the oxygen abundance in extragalactic \HII regions, similar to the N$_{2}$-method, but using the O$_{3}$N$_{2}$ $\equiv$ log[([O{\small{III}}] $\lambda$5007/H$\beta$)/([N{\small{II}}] $\lambda$6583/H$\alpha$)] index. This relation was revised by \citet{2004MNRAS.348L..59P} using 137 extragalactic \HII regions. They found that the relation is tight at O$_{3}$N$_{2}$ $\leq$ 1.9. The least square linear fit to the data in the range -1 $\textless$ O$_{3}$N$_{2}$ $\textless$ 1.9 yields the relation:   

\begin{equation}
12 + {\rm log(O/H)} = 8.73 - 0.32{\rm O_{3}N_{2}}
\end{equation}

In our sample of dwarf WR galaxies, all the \HII regions satisfied the condition O$_{3}$N$_{2}$ $\leq$ 1.9 that requires to use the above given relation. The oxygen abundances estimated from the O$_{3}$N$_{2}$-method are given in the fourth column of Table~\ref{tab:05}. 

We also used P-method to estimate the oxygen abundance. This method was proposed by \citet{2000A&A...362..325P,2001A&A...369..594P} and achieved a good agreement to the results obtained with the direct T$_{e}$-method. Pilyugin found that the precision of oxygen abundance determination with this method is $\sim$ 0.1 dex. This method uses the index R$_{23}$ and excitation parameter P, where these parameters are defined as:

\begin{table*}
\centering
\caption{The values of 12+log(O/H) determined from different indicators.}
\vspace {0.3cm}
\begin{tabular}{cccccccc}
\hline 
Galaxy & knot & N$_{2}$-Method & O$_{3}$N$_{2}$-Method & P-Method & T$_{e}$-Method & Median & Weighted mean \\
\hline
  && 12+log(O/H) & 12+log(O/H)  & 12+log(O/H) & 12+log(O/H)& 12+log(O/H) &12+log(O/H)\\
\hline
IC 3521      & \#a   & 8.61$\pm$0.01 & ---           & ---           & ---           & 8.61$\pm$0.01 & 8.61$\pm$0.01\\

IC 3521      & \#b   & 8.59$\pm$0.01 & ---           & ---           & ---           & 8.59$\pm$0.01 & 8.59$\pm$0.01\\

CGCG 038-051 & \#a   & 7.97$\pm$0.02 & 8.00$\pm$0.01 & 8.11$\pm$0.07 & 7.69$\pm$0.09 & 7.99$\pm$0.01 & 7.99$\pm$0.01\\

CGCG 038-051 & \#b   & 7.99$\pm$0.02 & 8.01$\pm$0.01 & 8.10$\pm$0.09 & 7.74$\pm$0.02 & 8.00$\pm$0.01 & 7.96$\pm$0.01\\

CGCG 038-051 & \#c   & ---           & ---           & ---           & 8.01$\pm$0.05 & 8.01$\pm$0.05 & 8.01$\pm$0.05\\

CGCG 041-023 & \#a   & 8.08$\pm$0.01 & 8.07$\pm$0.01 & 8.11$\pm$0.03 & 8.07$\pm$0.06 & 8.08$\pm$0.01 & 8.08$\pm$0.01\\

CGCG 041-023 & \#b   & 8.23$\pm$0.10 & ---           & ---           & ---           & 8.23$\pm$0.10 & 8.23$\pm$0.10\\

CGCG 041-023 & \#c   & 8.20$\pm$0.03 & ---           & ---           & ---           & 8.20$\pm$0.03 & 8.20$\pm$0.03\\

SBS 1222+614 & \#a+b & 7.93$\pm$0.02 & 7.93$\pm$0.01 & 8.10$\pm$0.03 & 8.06$\pm$0.02 & 8.00$\pm$0.02 & 7.96$\pm$0.01\\

SBS 1222+614 & \#c   & 7.98$\pm$0.02 & 7.96$\pm$0.01 & 8.12$\pm$0.08 & 7.98$\pm$0.01 & 7.98$\pm$0.02 & 7.97$\pm$0.01\\
\hline 
\end{tabular}
\label{tab:05}
\end{table*}

\begin{equation}
{\rm R_{23}} = \frac{{\rm [O{\small{II}}] \lambda 3727} + {\rm [O{\small{III}}] \lambda 4959} + {\rm [O{\small{III}}] \lambda 5007}}{{\rm H_{\beta}}}
\end{equation}

and

\begin{equation}
{\rm P} = \frac{{\rm [O III] \lambda 4959} + {\rm [O III] \lambda 5007}}{{\rm [O II] \lambda 3727} + {\rm [O III] \lambda 4959} + {\rm [O III] \lambda 5007}}
\end{equation}

This method uses two-zone models for \HII regions: a moderately high-metallicity \HII region with 12 + log(O/H) $\geq$ 8.2; and a low-metallicity \HII region. The best fit to the relations which can be adopted for the oxygen abundance determination for the two-zone models are given as: 

\begin{equation}
12 +{\rm log(O/H)_{high}} = \frac{{\rm R_{23}} + 54.2 + {\rm 59.45~P} + {\rm 7.31~P^{2}}}{6.07 + {\rm 6.71~P} + {\rm 0.37~P^{2}} + {\rm 0.243~R_{23}}}
\label{eq:06}
\end{equation}

and

\begin{equation}
12 + {\rm log(O/H)_{low}} = 6.35 + 1.45~{\rm log~R_{3}} - 3.19~{\rm log~P}
\label{eq:07}
\end{equation}

\noindent where the index R$_{3}$ is defined as ([O{\small{III}}] $\lambda$4959 + [O{\small{III}}] $\lambda$5007)/H$_{\beta}$. The oxygen abundance of all the spatially-resolved \HII regions in our sample of dwarf WR galaxies were estimated by taking the mean of the result from Eq.~\ref{eq:06} and~\ref{eq:07}, because the \HII regions in our sample seem to belong to the intermediate metallicity region. The estimated oxygen abundances using P-method are given in the fifth column of Table~\ref{tab:05}. It may be noted here that all these empirical relations to estimate oxygen abundance use emission line ratios for lines at closeby wavelengths, hence these methods are not very sensitive to errors in extinction corrections or flux calibration.

\section{Discussions}

\subsection{Comparison of metallicities from different indicators}

\begin{figure*}
\centering
\includegraphics[width=9.5cm,height=13.0cm,angle=270]{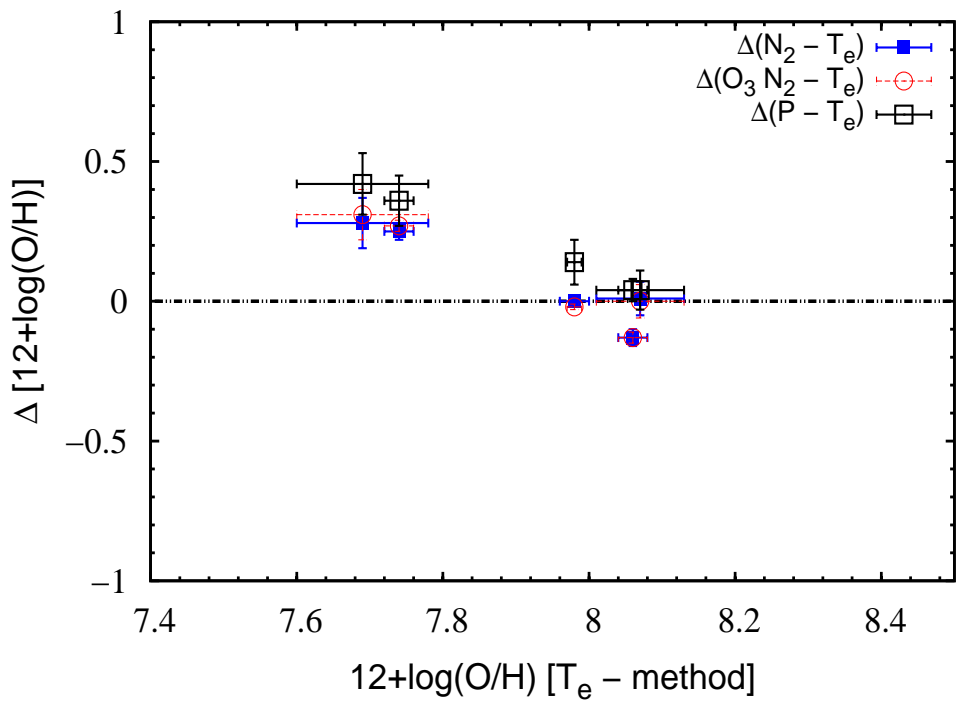}
\caption{The oxygen abundance difference between various empirical methods and the direct T$_{e}$-method. The abscissa denotes the oxygen abundance derived from the T$_{e}$-method.}
\label{fig:11}
\end{figure*}

The estimates of metallicity from different methods are given in Table~\ref{tab:05}. The metallicities derived from strong emission-line methods may give significantly biased results if the region under study have different structural properties (e.g., hardness of the ionizing radiation field and morphology of the nebulae) than those estimated using the empirically calibrated methods \citep{2010IAUS..262...93S}. Therefore, in order to test reliability of the empirical calibration relations, we made comparisons of the T$_{e}$-based metallicity with the metallicity derived using N$_{2}$-, O$_{3}$N$_{2}$- and P-method. In our comparison as shown in Fig.~\ref{fig:11}, all the empirical methods i.e., N$_{2}$-, O$_{3}$N$_{2}$- and P-method are found to yield metallicity very close to that estimated from the direct T$_{e}$-method. The best agreement is found between T$_{e}$- and N$_{2}$-methods, which shows a difference in the metallicities less than 0.06 dex on an average. In the literature, it is however noticed that the P-method shows the best agreement with T$_{e}$-method for high metallicity galaxies \citep[e.g.,][]{2005A&A...437..849S,2010A&A...517A..85L}, and such high metallicity galaxies are absent in our analysis. It can be seen that the two \HII regions which have low-metallicity (12+log(O/H) $\textless$ 7.8 show large differences up to $\sim$ 0.4 and $\sim$ 0.3 dex, respectively, in all the cases. The large difference for the two regions with low metallicity is not surprising as it is known that the empirical methods overestimate the oxygen abundance in comparison to that from the T$_{e}$-method in low-metallicity regions \citep{2000A&A...362..325P,2001A&A...369..594P}. Such large differences in low-metallicity regions were also noticed by \citet{2010A&A...517A..85L}. Overall, it can be stated that although there is a large scatter, the empirical methods provide estimates of metallicities in good agreement with those derived from the direct T$_{e}$-method. The difference between empirical and T$_{e}$-based estimates is between 0.02 to 0.4 dex. This estimated difference is in good agreement with those derived in other similar studies available in the literature \citep[e.g.,][]{2005A&A...437..849S,2008ApJ...681.1183K,2010A&A...517A..85L}. Overall, the presented analysis in this work is consistent with similar previous studies available in the literature \citep{2005A&A...437..849S,2007ApJ...670..457G,2009ApJ...700..654E,2010A&A...517A..85L}.

A detailed discussion on reasons responsible for differences in the metallicity estimates from different methods can be found in \citet{2003ApJ...591..801K} and \citet{2010A&A...517A..85L}. The observed differences between various estimates of metallicities are also discussed in \citet{2002astro.ph..7500S} and \citet{2005A&A...437..849S}. The observed discrepancies in the estimates of oxygen abundance from different methods are explained in terms of two main sources: (i) a lack of sufficient number of \HII regions with an accurate T$_{e}$-based estimates of oxygen abundance used in calibrations of the empirical methods, and (ii) a systematic offset between the observed forbidden-line temperatures and the nebular electron temperature used for calibrating photoionization models \citep{2003ApJ...591..801K}. The latter effect is more important due to the fact that the temperature fluctuations and gradients are known to exist in the ionized gas of star-bursting galaxies \citep{1967ApJ...150..825P,2002astro.ph..7500S,2005A&A...434..507S,2007ASPC..374...81P}. Moreover, \citet{2006ApJ...644..890G,2007A&A...464..885G} in their work aimed at determining the Balmer jump temperature in a large sample of low-metallicity \HII regions quantified the temperature fluctuation parameter to be $t^{2}$ $\leq$ 0.02 in the high temperature (T$_{e}$ $\textgreater$ 1.1 $\times$ 10$^{4}$ K) star-bursting regions. The differences can also be due to several other factors such as galaxies with different ionization parameters, different chemical evolution and star formation histories. Therefore, a proper choice of empirical methods for estimating metallicity plays an important role and must be used with caution.

\subsection{Spatial variations in chemical compositions} \label{mv}

The estimates of various chemical abundances such as O, N, Ne, S and Ar from the direct T$_{e}$-method and oxygen abundance from different indicators for the spatially-resolved star-forming region in the galaxies are given in Tables~\ref{tab:04} and~\ref{tab:05} respectively. In these tables, it can be noticed that the chemical abundances across the galaxies are, in general, homogeneous within the uncertainties in the estimates, except that in case of knot \#c in the galaxy CGCG 038-051 where the metallicity is found to be considerably different ($\sim$ 0.3 dex) from that found for other knots in the same galaxy. We show here that such differences can be due to uncertainty in the electron temperature which in these cases was assumed as 10$^{4}$ K in absence of direct T$_{e}$ estimate due to non-detection of temperature sensitive oxygen line of [O{\small{III}}] $\lambda$4363. Based on an investigation of chemical history of dwarf galaxies studied in the literature which show homogeneous chemical abundances, we found that such galaxies usually show a nearly constant electron temperature in all distinct \HII regions within the same galaxies \citep{1997ApJ...477..679K,2009AJ....137.5068L}. The constant electron temperature is also seen in the cases of different knots in SBS 1222+614 and CGCG 038-051 in the present work with measured T$_{e}$. In CGCG 038-051, temperature for the knot \#c was assumed as 10$^{4}$ K in Table~\ref{tab:04}, while temperature for other knots have T$_{e}$ close to $\sim$ 16,500 K. We therefore re-estimated the oxygen abundance for the knot \#c in CGCG 038-051 by assuming the same electron temperature as those estimated for other knots in the galaxies (see Table~\ref{tab:04}). The revised value for metallicity for the knot \#c is found as 7.40$\pm$0.12 in CGCG 038-051. This value is nearly close to those estimated for other knots in the same galaxy within the measurement uncertainties. Our analysis implies that an uncertainty in electron temperature can lead to a considerable difference in the metallicity estimates for distinct \HII regions. This also implies that uncertainty in the estimates of electron temperature plays a major role in the estimates of oxygen abundance. \citet{2006A&A...459...71I} observed several \HII complexes in SBS 0335-052E and found a decreasing trend of oxygen abundance between $0.10 - 0.14$ dex as they proceeded from one end to other end of the galaxy. This trend was interpreted as self-enrichment by heavy elements. However, they also suggested that the error in the estimates of T$_{e}$ can lead to an apparent variation in the oxygen abundance in SBS 0335-052E. Overall, we conclude that all four galaxies in our sample are chemically homogeneous, similar to other galaxies studied in the literature \citep[e.g.,][]{1989ApJ...347..875S,1996ApJ...471..211K,1998ApJ...497..601K,2006ApJ...642..813L,2009ApJ...705..723C,2012ApJ...754...98B}. 

The separations between the \HII regions in these galaxies are in the range of $0.3-2.4$ kpc. This implies that a homogeneity in the chemical abundances in the studied galaxies is observed over large spatial scales. Such a chemical homogeneity in galaxies can be expected if the processed metals injected into the ISM from past bursts of star formation have cooled down, well mixed and homogeneously distributed across the whole extent of the galaxy. While the newly synthesized metals formed in the current episode of star formation are still in hot gas-phase \citep[T$\sim$10$^{7}$ K;][]{1996ApJ...471..211K,2006A&A...454..119P,2009ApJ...707.1676C,2009AJ....137.5068L,
2011MNRAS.412..675P,2011MNRAS.414..272H,2012MNRAS.423..406G,2013AdAst2013E..20L}, not observable at optical bands. Therefore, the currently observed metals in the optical band are from previous episodes of star formation. A possible mechanism responsible for the metal dispersal and mixing at large spatial scales can be bar-induced rotation or shear, in particular, in massive galaxies \citep[e.g.,][]{1995A&A...294..432R}. Alternatively, the global hydrodynamical process such as starburst-driven super-shells and/or gas inflows can be another possible mechanism for transporting and mixing the metals over the whole extent of the galaxy in typical timescales of few 10$^{8}$ yr \citep{1996AJ....111.1641T}.

A metal enrichment to the ISM local to the star-bursting region can take place due to winds from the most massive stars and the supernova explosions near the youngest star-bursting region. For example, \citet{1996ApJ...471..211K} reported an oxygen overabundance by $\sim$ 0.1 dex at the locations of the young starburst in NCG 4214 (dwarf irregular WR galaxy), possibly from recent supernova events. At least one of star-forming regions ($\leq$ 6 Myr in age) in each galaxy studied in the present work are observed in WR phase, which appears before supernovae explosions of massive stars \citep{2005A&A...429..581M}. Therefore, there is a possibility that the ejection of newly synthesised heavy metals in such star-forming regions in WR phase have not yet taken place through supernova explosions. In fact, \citet{2016MNRAS.462...92J} have predicted based on radio continuum data analysis that these galaxies have radio deficiency, most likely due to a lack of recent supernova events. However, Nitrogen-enrichment as expected in these star-forming regions is not detected. The metal enrichment of local ISM through ejection in the luminosity-driven stellar winds from massive WR and/or other O or OB-type stars will be too small to be detected in the present work. Based on these analysis, we have concluded that the observed chemical homogeneity in our sample of dwarf WR galaxies hosting very young massive star formation ($\leq$ 10 Myr) is most likely a consequence of the presence of cooled and well mixed metals formed in the previous episodes of star formation ($\textgreater$ 100 Myr) and the metals formed in the current episode of star formation are still likely to be in hot gas-phase and not seen in optical bands.

\subsection{$\alpha$-elements to oxygen ratios}

\begin{figure*}
\centering
\includegraphics[width=9.5cm,height=13.0cm,angle=270]{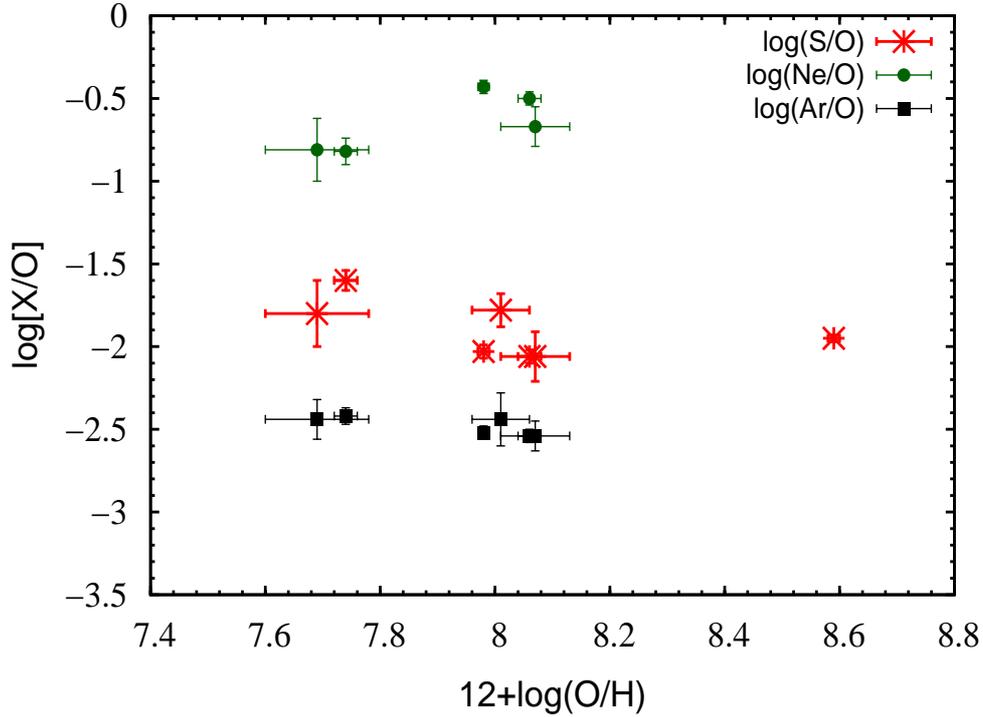}
\caption{The values of log(Ne/O), log(S/O) and log(Ar/O) plotted against 12+log(O/H), as determined in this work for our sample of dwarf WR galaxies.}
\label{fig:12}
\end{figure*}

Fig.~\ref{fig:12} shows the values of log(Ne/O), log(S/O) and log(Ar/O) as a function of 12+log(O/H) for all the spatially-resolved \HII regions analyzed in this work. We find that the abundance ratios of S, Ne and Ar relative to oxygen are nearly constant, independent of metallicity. These trends are consistent with those seen in other dwarf galaxies previously studied in the literature \citep{1999ApJ...511..639I,2006A&A...448..955I,2010A&A...517A..85L}. From the present work, the mean values of log(Ne/O), log(S/O) and log(Ar/O) ratios are estimated as -0.65$\pm$0.05, -1.86$\pm$0.04 and -2.48$\pm$0.04, respectively. These values are comparable within errors with the previously reported values for star-bursting dwarf galaxies \citep[e.g.,][]{1999ApJ...511..639I,2006A&A...448..955I}. \citet{1999ApJ...511..639I} estimated the mean values of log(Ne/O), log(S/O) and log(Ar/O) ratios as -0.72$\pm$0.06, -1.56$\pm$0.06 and -2.26$\pm$0.09 respectively. Moreover, \citet{2010A&A...517A..85L} found the mean values as 0.70$\pm$0.13, -1.68$\pm$0.10 and -2.37$\pm$0.12 for the log(Ne/O), log(S/O) and log(Ar/O) ratios, respectively, in a similar sample of dwarf WR galaxies as studied in the present work. 

Despite of uncertainties in the estimated values of $\alpha$-elements to oxygen ratios, we noticed that the Ne/O and S/O ratios show slightly increasing and decreasing trends with increasing metallicity respectively. This trend has also been reported in the literature. For example, \citet{2006A&A...448..955I} reported an increasing trend of Ne/O ratio with increasing oxygen abundance, most likely due to a moderate depletion of oxygen into dust grains in metal-rich galaxies. Similarly, \citet{2003A&A...403..829V} reported a slight decreasing trend in S/O ratio in relatively high-metallicity star-burst galaxies due to depletion of sulfur onto dust grains. 

\begin{figure*}
\centering
\includegraphics[width=9.5cm,height=13.0cm,angle=270]{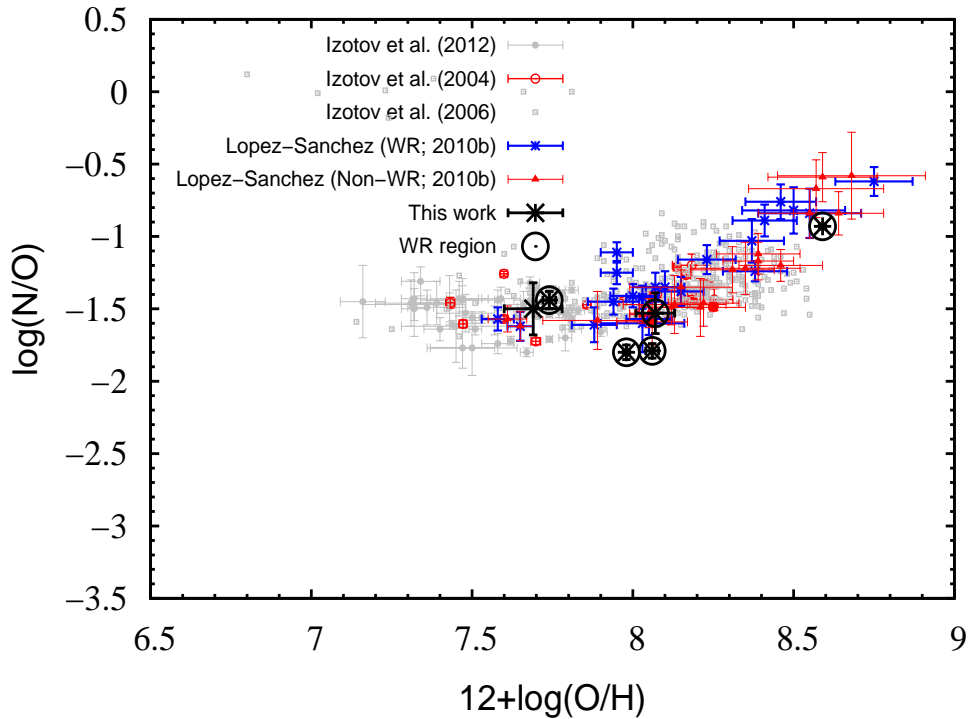}
\caption{log(N/O) vs 12+log(O/H), as determined from this study for our sample of dwarf WR galaxies and for other galaxies from the literature.}
\label{fig:13}
\end{figure*}

In a scenario of chemical evolution in galaxies, it is believed that the N/O ratio is a powerful indicator of galaxy evolution \citep{2004A&A...425..849P,2006MNRAS.372.1069M}. The origin of nitrogen in galaxies is an ongoing debate since several decades. In Fig.~\ref{fig:13}, we investigated the relation between log(N/O) ratio and oxygen abundance [12+log(O/H)] for a large sample of dwarf WR galaxies with our data also included in it. In this figure, it can be seen that the values of log(N/O) show a varying trend depending on two metallicitiy regimes: low-metallicity (12+log(O/H) $\lesssim$ 8) and high-metallicity (12+log(O/H) $\gtrsim$ 8). The low-metallicity \HII regions have a nearly constant value of log(N/O) ratio independent of 12+log(O/H), while an increasing trend for log(N/O) ratio with increasing oxygen abundance is observed in the high-metalicity \HII regions. This trend is well known and have been reported in various studies. However, it is still not completely understood and is presently explained in the literature as follows: the observed behavior of log(N/O) with 12+log(O/H) is generally explained in terms of two sources of enrichment termed as primary and secondary. Massive stars produce small amounts of nitrogen in early phase of evolution, which is termed as the primary production of nitrogen \citep[]{1978MNRAS.185P..77E,1979A&A....78..200A,1999ApJ...511..639I}. The low and intermediate mass stars produce nitrogen and other elements heavily enriching ISM with a time lag as compared to the primary production. The latter delayed process is often termed as the secondary production \citep{2000ApJ...541..660H}. The low metallicity region (12+log(O/H) $\lesssim$ 8) with a constant N/O ratio is believed to be from the primary production of nitrogen in massive stars \citep[e.g.,][]{1990PASP..102..230G,1993MNRAS.265..199V,1995ApJ...445..108T,1998ApJ...497L...1V}. At high metallicity, a steep increase in N/O ratio is believed to be due to increased secondary production and partly also due to selective depletion of oxygen in dust grains \citep[e.g.,][]{1998ApJ...497..601K,1999ApJ...511..639I,2003A&A...397..487P,2006MNRAS.372.1069M,2006A&A...448..955I}.

Some WR galaxies also show excess nitrogen in the range of $0.25 - 0.85$ dex in the relation between log(N/O) and 12+log(O/H) over all the metallicity regimes \citep{1997ApJ...477..679K,2004A&A...419..469P,2007ApJ...656..168L,2008A&A...485..657B,2010A&A...517A..85L,2013JApA...34..247J,
2014MNRAS.439..157K}. The excess of nitrogen in WR galaxies is attributed to nitrogen ejection in luminosity-driven stellar winds of WR stars. However, we did not detect excess nitrogen in these star-bursting regions with WR features in our galaxy sample (see Fig.~\ref{fig:13}). The N/O ratio is found to be consistent with the normal trend known for non-WR star-bursting galaxies. Presently, this behaviour is not completely understood. Consistent with our previous conclusion as made in Sect.~\ref{mv}, it is possible that the ejected extra nitrogen from WR stars is not yet sufficiently cooled down to be detected at optical wavebands. It is worth to point out that absence of extra nitrogen has also been reported in several other WR galaxies \citep[e.g.,][]{1996ApJ...471..211K,2012A&A...544A..60M,2013MNRAS.430.2097J,2012MNRAS.424..416W}. 

\subsection{Luminosity-metallicity relation} 

\begin{figure*}
\centering
\includegraphics[width=9.5cm,height=13.0cm,angle=270]{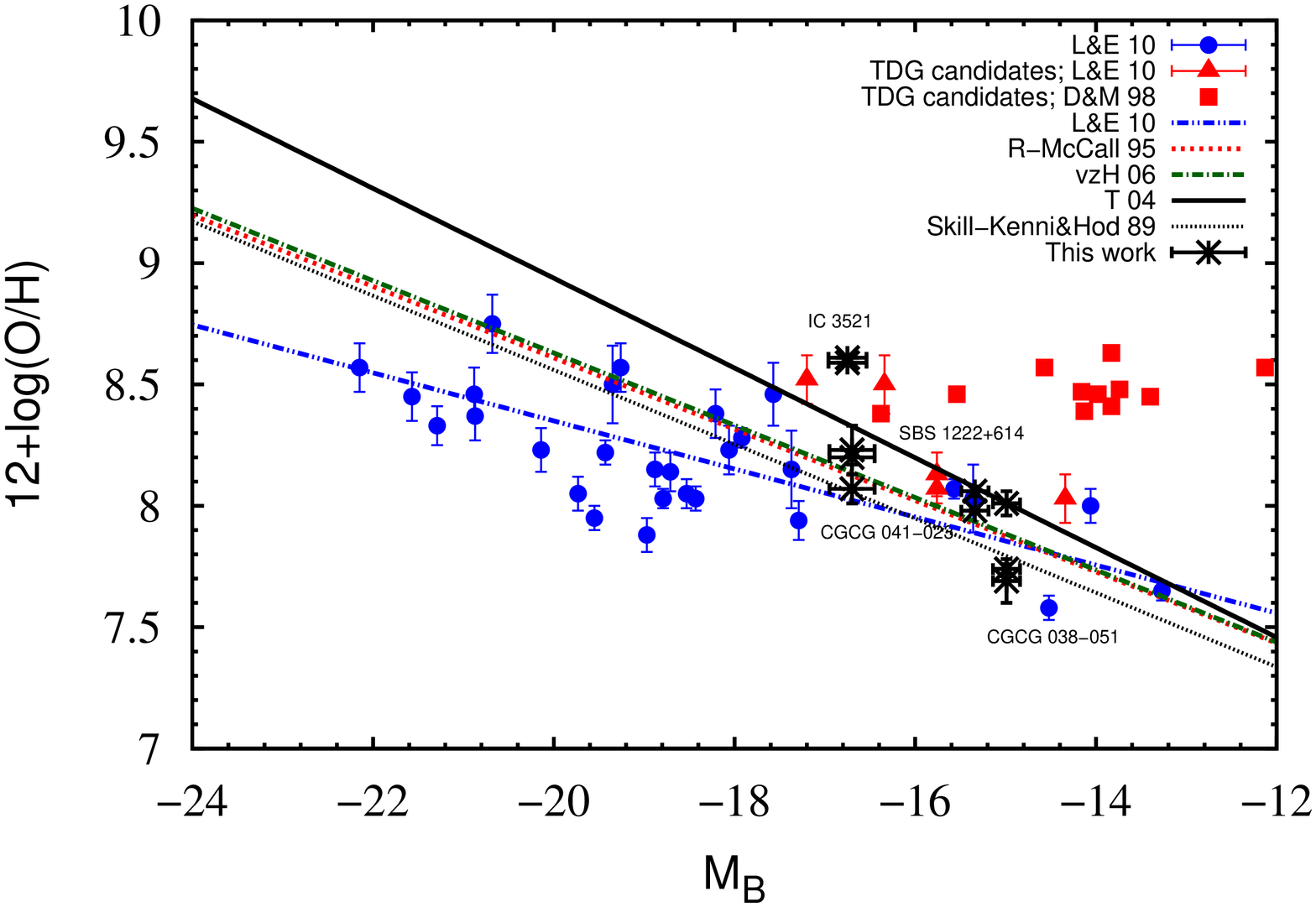}
\caption{The luminosity-metallicity relation for the galaxies studied in this work and in the literature. The metallicity is expressed in units of 12+log(O/H) and the luminosity is expressed in terms of absolute magnitude in B-band.}
\label{fig:14}
\end{figure*}             

The luminosity-metallicity relation for galaxies has been used as a tool for tracing evolution of galaxies as more luminous galaxies are supposed to contain a large fraction of processed material \citep{1997ApJ...481..689M,2000MNRAS.312..497B,2001AJ....121..753B}. This relation is known since the early works of \citet{1979A&A....80..155L} and \citet{1984ApJ...281L..21R}. Thereafter, the relation was confirmed in many similar studies in the literature \citep[e.g.,][]{1984AJ.....89.1300B,1985ApJ...296L...1W,1989ApJ...347..875S,1992MNRAS.259..121V,1994ApJ...420...87Z,
2002ApJ...581.1019G,2005A&A...437..849S,2010A&A...517A..85L,2004ApJ...613..898T,2006ApJ...636..214V,1998A&A...333..813D,
1995ApJ...445..642R}. We plotted the B-band luminosity-metallicity relation for galaxies in our sample in Fig.~\ref{fig:14}. In this figure, we have included other similar dwarf WR galaxies from the sample of \citet{2010A&A...517A..85L} and their best linear-fit relations. This figure also includes some other similar relations reported in the literature, e.g., the known luminosity-metallicity relation for dwarf and irregular galaxies in \citet{1995ApJ...445..642R}, \citet{2006ApJ...636..214V}, \citet{2004ApJ...613..898T} and \citet{1989ApJ...347..875S}. Here, our galaxies appear to follow normal luminosity-metallicity relation. The luminosity-metallicity relation has also been used to identify tidal dwarf galaxies by locating high metallicity galaxies for their given optical luminosity \citep{1998A&A...333..813D,2010A&A...517A..85L}. Our analysis indicates that the dwarf WR galaxies studied in the present work trace a normal evolution as seen for other normal dwarf galaxies and these are not tidal dwarf galaxies.

\subsection{Local galaxy environment}  

In this section, we discuss the local environment of each individual galaxies in our sample. The galaxy density within a comoving volume of 0.5 Mpc in projected radius and $\pm$ 250 km~s$^{-1}$ in radial velocity range from the recession velocity of the target galaxies is presented. The galaxy density is estimated using the NED tool (\textit{https://ned.ipac.caltech.edu/forms/denv.html}). The local galaxy environment of each individual galaxies are discussed below.

\subsubsection{IC 3521}

IC 3521 belongs to the Virgo Cluster of galaxies. A total of 41 galaxies are listed in the NED in the defined space volume, implying an average galaxy density of $\sim$ 80 Mpc$^{-3}$. All the neighbour galaxies have velocities in the range of $439-833$ km~s$^{-1}$. This indicates the presence of a galaxy-rich dense environment around IC 3521. Among these neighbour galaxies, a total of nine and three galaxies in the vicinity of IC 3521 are dE and dIrr types, respectively, and only nine galaxies are big spiral or lenticular types. A total of twenty one galaxies are still morphologically unclassified.     

\subsubsection{CGCG 038-051}

In the defined volume, a total of 9 galaxies are listed. It represents an average galaxy density of $\sim$ 19 Mpc$^{-3}$. The neighbour galaxies are found to be in a narrow velocity range of $986-1070$ km~s$^{-1}$. Except two dwarf galaxies (LSBC L1-137 and LSBC L1-137A), all other galaxies are giant systems which have well-developed bright spiral arms and disk. These galaxies together appear to be residing in a group-like environment.    

\subsubsection{CGCG 041-023}

CGCG 041-023 belongs to the VV 462 galaxy group. In the vicinity of CGCG 041-023, a total of 3 galaxies are listed in the NED which in turn gives an estimate of galaxy density as $\sim$ 8 Mpc$^{-3}$ within the defined volume. Two neighbour galaxies have velocity in a very narrow range of $1350-1359$ km~s$^{-1}$, while one has velocity about $\sim$ 1482 km~s$^{-1}$. Interestingly, all these neighbour galaxies are dwarf systems, and their SDSS colour composite images show presence of blue regions extended over the galaxies extent representing an ongoing star formation activities in them. It seems that CGCG 041-023 together with neighbour galaxies forms a small group of star-forming dwarf galaxies. 

\subsubsection{SBS 1222+614} 

Within the defined volume, a total of 4 galaxies are listed, implying an average galaxy density of $\sim$ 10 Mpc$^{-3}$. All these neighbour galaxies have velocities in a very narrow range of $709-722$ km~s$^{-1}$, except for SDSS J12195.11+613110.0 which has velocity of 516 km~s$^{-1}$. Interestingly, all the neighbour galaxies are dwarf galaxies which are blue in optical band as seen in their SDSS colour composite images. The galaxies MCG +10-18-044 and UGC 7534 are closely associated with SBS 1222+614 and UGC 7544 as revealed in the work of \textcolor{blue}{Jaiswal \& Omar (private communication)} and \citet{2002A&A...389...29S,2002A&A...389...42S}. SBS 1222+614 together with its neighbour galaxies indicates a small group of galaxies.

\section{Summary and conclusions}

The spatially-resolved optical spectroscopic observations of four nearby dwarf Wolf-Rayet (WR) galaxies were presented here. These galaxies are residing in group and cluster environments with widely varying galaxy density in the range of $8-80$ Mpc$^{-3}$. This environment is suitable for galaxy-galaxy interactions, which might have triggered star formation in these galaxies. The ages of the most recent starburst events in the galaxies are found between 3 and 10 Myr. The gas-phase metallicity [12+log(O/H)] for all the spatially-resolved star-forming regions is derived using several indicators and compared with each other. This comparison indicated that although there is a large scatter in the estimates of metallicities from different indicators, the empirical methods provide the estimates of metallicities which are in good agreement with those derived from the direct T$_{e}$-method. Consistent with other similar studies available in the literature, the differences between empirical and T$_{e}$-based estimates of oxygen abundances are found between 0.02 to 0.4 dex. This study also shows that different star-forming regions within the galaxies are chemically homogeneous. Against an expectation of N-enrichment in WR galaxies, these galaxies show a normal N/O ratio for their given metallicities. It is speculated here that the newly synthesized metals from the current episode of star formation in these WR galaxies are possibly in hot gas-phase, and the metals from the previous episodes have cooled down and well mixed across the whole extent of galaxies, which makes galaxies chemically homogeneous with normal N/O ratio. The luminosity-metallicity relation for these galaxies are consistent with the previously known relation for normal dwarf and large spiral galaxies, indicating that these dwarf WR galaxies are evolving in normal way and do not belong to a category of tidal dwarf galaxies.


\section*{Acknowledgements}

We thank the anonymous referee for his valuable comments which improved the contents and quality of the paper. This research has made use of the NASA/IPAC Extragalactic Data base (NED) and the Smithsonian Astrophysical Observatory (SAO)/NASA Astrophysics Data System (ADS) operated by the SAO under a NASA grant. We thank the staff of the HCT who made the observations possible. The HCT is operated by Indian Institute of Astrophysics (IIA) through dedicated satellite communication from the Center for Research \& Education in Science \& Technology (CREST), IIA, Hosakote, Bangalore, India. 

\bibliographystyle{mnras} 
\bibliography{references}
\label{lastpage}
\end{document}